\documentclass[twocolumn]{29onr_art}
\usepackage{natbib}
\usepackage{ulem}
\usepackage{graphicx}
\usepackage{fixltx2e}
\usepackage{amsmath,float}
\usepackage{graphicx}
\usepackage[english]{babel}
\usepackage{epsfig}
\usepackage{color}
\usepackage[colorlinks=true,
            urlcolor=black,
            linkcolor=black,
            citecolor=black,
            bookmarks=true,
            pdftitle={The Impact of a Deep-Water Plunging Breaker},
            pdfauthor={An Wang},
            pdfsubject={Abstract Submitted for the 31st Naval Hydrodynamics Symposium},
            pdfkeywords={Breaking Waves, Wave Impact, Air Entrainment, Numerical Simulations},
	    bookmarksopen=false,
            pdfpagemode=None]{hyperref}

\definecolor{docgreen}{rgb}{0,.5,0}

\pdfoutput = 1

\begin{document}

\title{Spray Formation during the Impact of a Flat Plate on a Water Surface}

\author{
An Wang$^1$, Shizhao Wang$^2$,  Elias Balaras$^2$\\
Devin Conroy$^3$, Thomas T. O'Shea$^3$ and James H. Duncan$^1$}

\affiliation{\small $^1$University of Maryland, College Park, Maryland, U. S. A. \\
$^2$George Washington University, Washington D. C., U.S.A. \\
$^3$Leidos, La Jolla, California, U.S.A.}

\maketitle

\begin{abstract}
In naval hydrodynamics most the reseach on the impact of the hull of a
vesel to the water surface has focused primarily on mapping the forces
on the hull and much less to the formation of the spray.  The latter
is the topic of the present work, where the spray generated by the
impact of a flat plate on a quiescent water surface is studied by
experiments and simulations at different impact Froude numbers.
Overall two types of sprays are formed: Type I, which is a cloud of
droplets and ligaments generated at the impact of the plate's leading
edge with the water surface, and Type II, which is formed after the trailing
edge of the plate enters the local water surface.  Detailed analysis
of the experimental data for the Type II spray revealed that the spray
envelopes and root point trajectories are independent of Froude number
close to the trailing edge, while further away reach higher vertical
positions for larger Froude numbers. The numerical simulation captures
the general behavior of the Type II spray and agrees favorably with
the experimental results.  A series of computations with a flexible
plate is also reported, where it is shown that increased flexibility
reduces the height of the spray and suppresses the formation of
droplets.

\end{abstract}

\section{Introduction}
When planing vessels travel in rough sea conditions, the impact between  the hull and the water surface is an important issue that requires further understanding. During the past decades, the pressures and forces on the hull have been studied extensively; however, the formation of the spray during the impact has drawn less attention. The spray carries a significant amount of kinetic energy and part of the spray can easily get above the deck. The formation of the spray causes changes in the topology of the air-water interface and in some cases, air bubbles are entrained. Furthermore, the cloud of spray significantly increases the visibility of the vessel. Another important motivation for studying the spray formation is safety concerns during the landing of a seaplane on a water surface. The spray generated during this impact situation can impinge on the wings and cause significant structural damage \citep{savitsky1958main}. So far, little knowledge about the formation of  spray during the impact of a hull on water surface has been obtained.

In order to explore the physics of hull/water-surface impact, the problem is usually simplified to the impact of a vertically moving wedge with a quiescent water surface. There have been many theoretical studies of this problem. In most of these studies, potential flow is assumed since viscosity does not have sufficient time to affect the bulk flow due to the short time scale of the impact process and since, when no air is entrained, the water can be treated as incompressible.   
In one of the earliest of these theoretical works, \cite{von1929impact} applied conservation of momentum to predict the impact load. This model did not consider the local water rising along the wedge. \cite{wagner1932stoss} proposed a potential flow model that assumes blunt body impact and considers the local water rise, although this model overestimates the wetted area along the wedge. \cite{dobrovol1969some} derived a similarity solution based on \cite{wagner1932stoss}'s method and applied it to the water entry problem of a symmetric wedge. \cite{howison1991incompressible} extended Wagner's method and applied it to several 2D and 3D impact problems with different geometries of the impacting body, whose bottom was nearly parallel to the undisturbed water surface. Additional  theoretical studies of wedge slamming include \cite{faltinsen2008nonlinear} and \cite{moore2012three}. A review has been given by \cite{korobkin1988initial}.

Some early experimental work on wedge impact was reported in \cite{chuang1971drop}, \cite{chuang1973slamming}, \cite{greenhow1983nonlinear}. \cite{chuang1973slamming} reported a series of experiments of slamming a wedge model with combined horizontal and free falling motions. Both rigid and elastic models with various deadrise angles were used  and the deflection, pressure distribution and acceleration of the model were measured. These early experiments did not have sufficient resolution to study the detailed geometry of the spray. 

In several more recent studies, the spray generated during impact also was examined.   \cite{peters2013splash} performed an experimental and numerical study of the splash generated by the impact of a circular disk on a quiescent water surface with the disk oriented horizontally. It was found that the length and velocity scales follow a power law in time and that the scaled profiles of the splash at different times collapse to a single curve. \cite{marston2014ejecta} studied the impact of a solid cone with various angles into different liquids. The shape of the ejecta was observed to be self-similar at all speeds for low surface tension liquids or at high impact speeds for high surface tension liquids. In these axisymmetric experiments, the angle between the cone surface and the still water surface was relatively large. \cite{iafrati2004initial} did a numerical study on the vertical impact of a flat plate on water surface. It was found that the initial water surface profiles were self-similar at leading order. In addition, it was predicted that the thickness of the spray jet  decays as O($R^{-5}$) where $R$ is the distance measured from the plate edge. Surface tension was not included in these simulations. 

Droplets can break up from the continuous spray sheet when the sheet becomes thin and therefore unstable. \cite{peters2013splash} pointed out that droplets break up from the rim of the splash because of Rayleigh--Taylor instability. The diameter of the droplets are found to be nearly independent of the impact Weber number. 
 \cite{riboux2015diameters} studied drop impact on a solid surface and, in particular, the tiny droplets that are generated from the rim of the splash formed during the impact.  During the process, surface tension tries to slow down the rim of the splash, which is moving outward with high speed due to the momentum carried by the impinging droplet.  Due to instability, ridges form at the edge of the rim during the deceleration process. These ridges break up into droplets when their amplitudes become sufficiently large. The stability of the rim was studied in detail by \cite{krechetnikov2010stability}.

During the impact of a flat plate or a wedge on a water surface, air can be entrained in several scenarios. When the deadrise angle of the wedge is small or in the case of zero deadrise angle, the water surface can rise at the  edge of the plate and trap an air packet \citep{faltinsen2005hydrodynamics}.  \cite{semenov2009onset} reported, based on numerical calculations, that a negative pressure region can occur along the wedge in cases with oblique impact velocity.  This low pressure region might cause ventilation, cavitation or flow separation. Negative pressures were also found by \cite{reinhard2013water} in a study of the oblique impact of an elastic plate.  The authors indicated  that  regions of negative pressure and cavitation are more likely to happen for elastic plate impact because of the vibration of the plate. It was found that the entrained air  changed the behavior of the impact pressure significantly. \cite{iafrati2008hydrodynamic} suggested that the existence of entrained air can reduce the peak pressure but increase the loading period, which results in a large pressure impulse. 

%
%
In this paper, the vertical impact of a flat plate on a quiescent water surface is studied by experiments and both viscous and inviscid numerical simulations.   The spray generation for a single dead-rise (roll) angle and a range of impact velocities are studied.

\section{Methods}

\subsection{Experimental Setup}

\begin{figure}[!htb]
\begin{center}
\includegraphics[width=0.95\linewidth]{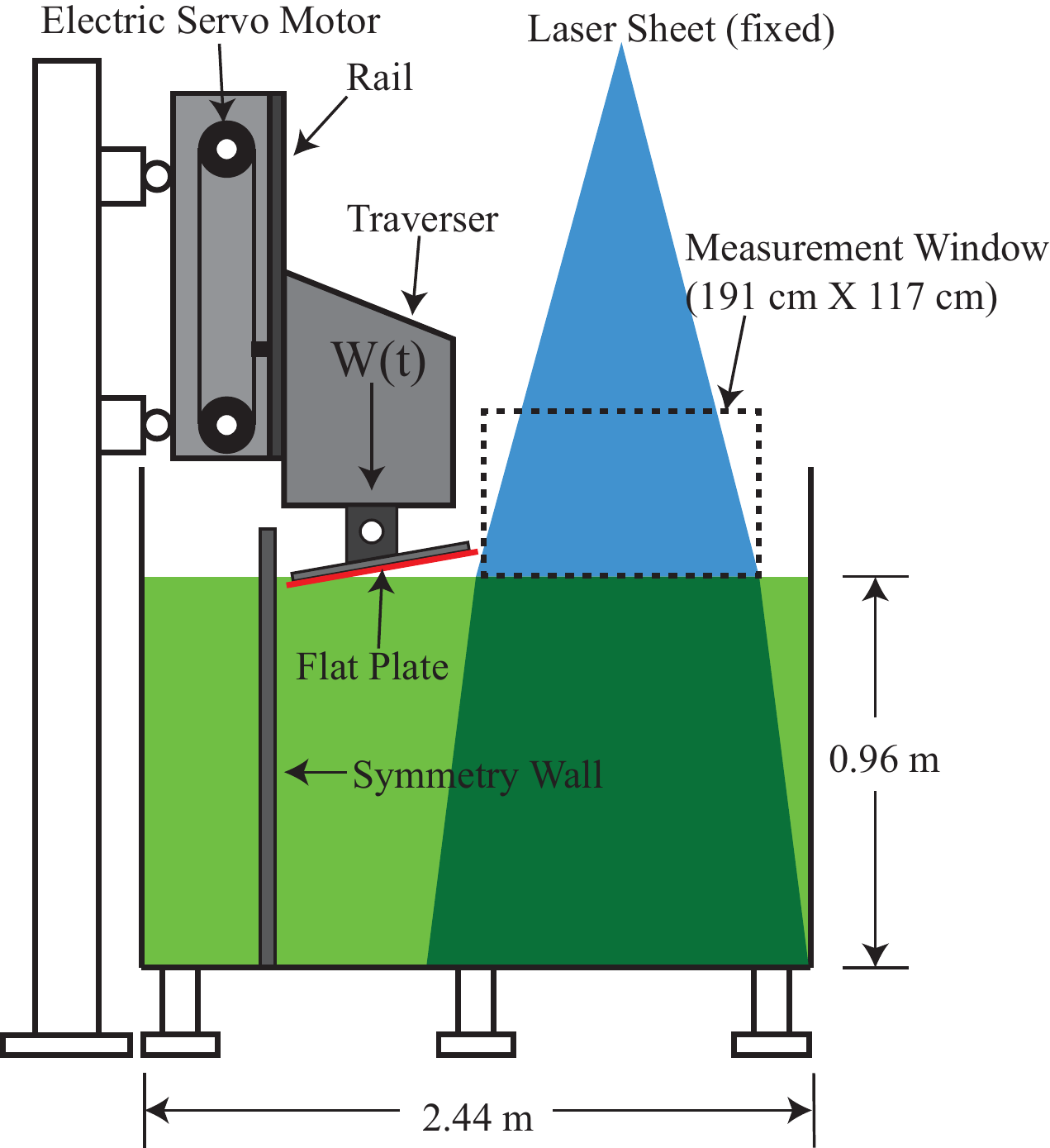}   \\
\end{center}
\vspace*{-0.15in}
\caption{The experimental facility and setup. \label{fig:setup}}
\end{figure}

\begin{figure}[!htb]
\begin{center}
 \includegraphics[width=0.95\linewidth]{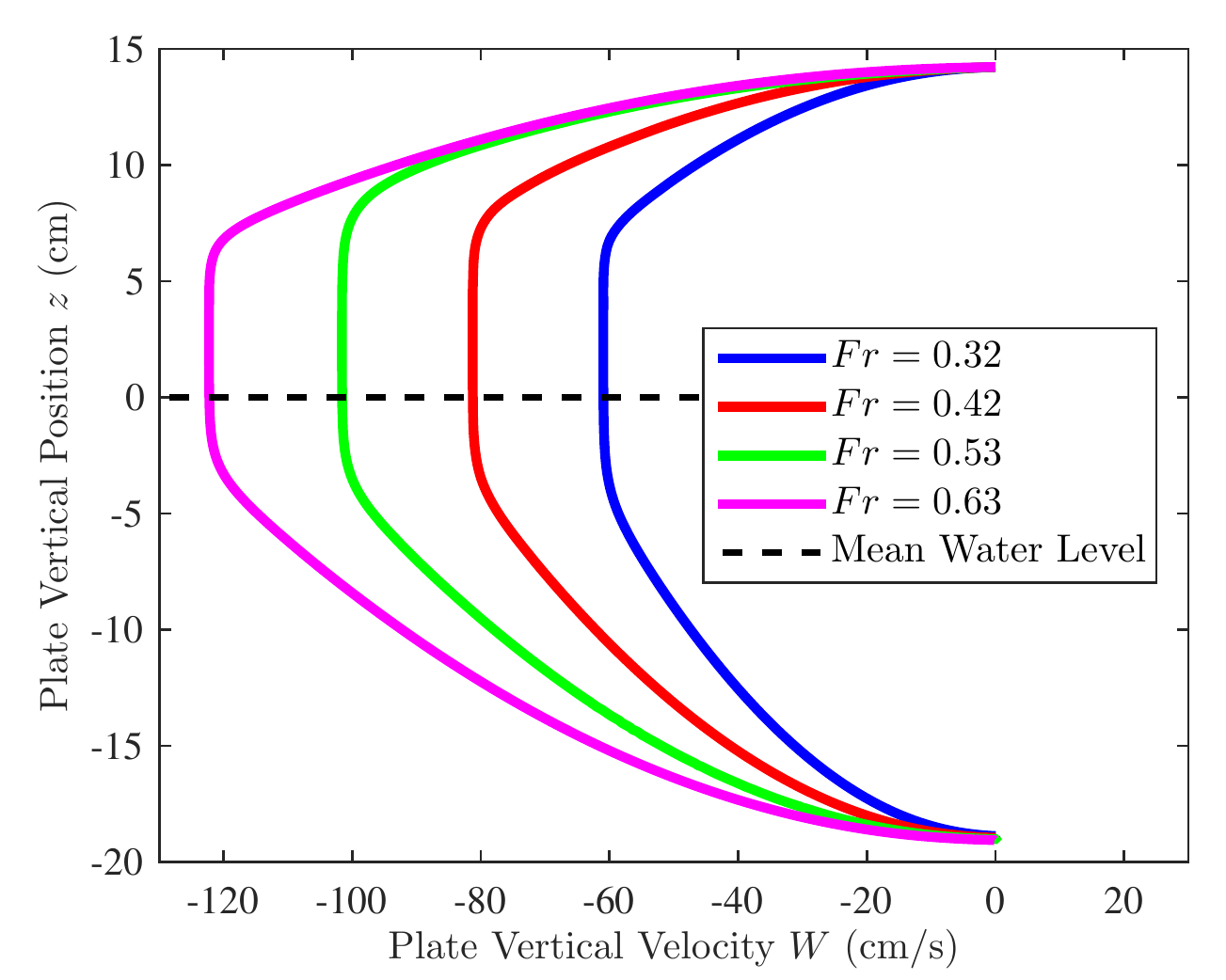}
 \end{center}
 \vspace*{-0.25in}
  \caption{Motion of the vertical traverser. The vertical axis is the vertical position of the leading edge of the plate. $Z=0$ is at the mean water level. Four different impact velocities are used. The Froude number corresponding to the vertical initial impact velocity $W_0$ is defined as $Fr=W_0/\sqrt{gB}$, where $B$ is the width of the plate.\label{fig:motion}}
\end{figure}

The experiments were performed in a towing tank that is 13.7~m long, 2.4~m wide and 1.3~m deep, see Figure~\ref{fig:setup}.  The tank includes a carriage that moves along the 13.7-m length of the tank.  This horizontal carriage is driven by a 60-horsepower hydraulic system and can achieve $3g$ accelerations and a maximum speed of 10~m/s. The model is attached to the horizontal carriage via a second carriage that moves vertically over a distance of 0.6~m. This vertical carriage is driven by an electric servo motor and has a maximum speed of about 2.5~m/s.  Both the horizontal and vertical motions are controlled by computer-based feedback systems that employ position sensors for both carriages.  The design of the tank has two additional features that are essential for the present experiments.  First, the entire carriage system is supported by 13.4-m-long rails that guide the horizontal motion of the carriage system and these rails are supported by a row of columns that are completely disconnected from the water tank, see Figure~\ref{fig:setup}, and mounted on the 0.86-m-thick steel-reinforced concrete floor of the laboratory.  This isolation between the carriage system and the tank prevents the carriage motion from causing tank wall vibrations, which would generate unwanted water waves and acoustic waves in the tank.   The second important feature of the tank is that since the carriage system is supported from only one side of the tank, there is an unobstructed view of the model and spray generation process from the ends and opposite side of the tank, with about 1.8~m of water surface to the opposite long tank wall.  This unobstructed view greatly facilitates the measurements of spray with optical-based systems.

In the present experiments, a flat polycarbonate plate with dimensions 1.22~m long by 0.38~m wide and 1.27~cm thick is mounted on a sturdy aluminum frame which is in turn attached to the vertical carriage.  For all experimental conditions, the horizontal carriage is held stationary.  The plate is rolled $10^\circ$ about its long axis so that the high edge of the plate is about 1.8~m from the far long side wall of the tank and both the high and low edges of the plate are parallel to the quiescent water surface.  The low edge of the plate is positioned about 2~mm from a symmetry wall that is placed on the interior of the tank.  The symmetry wall makes the single plate equivalent to a V-shaped wedge.

A cinematic laser-induced fluorescence (LIF) technique was used to measure the surface profiles of the spray. 
In this method, illumination is provided by a vertical light sheet created from a 7-Watt argon-ion laser. The plane of the light sheet is oriented perpendicular to the long axis of the plate and located at $0.75L$ from one edge of the plate, where $L=1.22$~m is the plate's length.
The water was mixed with fluorescent dye (Fluorescein Sodium Salt) during the experiments. Three high-speed cameras (25.6~mm $\times$ 16.0~mm sensor dimension, 2560 $\times$ 1600 pixel resolution, Vision Research) were installed at one  end of the towing tank and oriented with a slight downward angle.  The cameras were focused on the light sheet at a height just above the water surface.   Long-wavelength-pass optical filters were installed in front of each camera lens to block out specular reflections of the laser light but allow light from the fluorescing dye to reach the camera sensor. The cameras were set to record movies of the spray generation at a frame rate of  800~Hz. The motion of the traverser and the cameras were synchronized through a pulse-delay generator (Berkeley Nucleonics Corporation).

The images are slightly distorted due to lens imperfections and because the lines of sight of the cameras were not perpendicular to the plane of light sheet.  This distortion was correct via calibration images and image processing.  In this method, a calibration board with printed with black and white 2.54-cm checkerboard pattern was installed in the plane of the light sheet.  An image of this calibration board was recorded by all cameras. A numerical method was then used to correct for the image distortion.  To ensure the consistency of the calibration, images were taken and processed both before and after the spray measurements.  The water surface profiles were extracted from the LIF image based on the gradient of the image intensity. 

To generate the motion of the vertical traverser, an acceleration vs time curve is first created. The accelerations for the accelerating and decelerating period are a positive and negative constant for most of the period, respectively. The starting and ending portions of the two acceleration periods were smoothed with a hyperbolic tangent function to avoid the sudden changes in the temporal derivative of the acceleration.  The resulting acceleration vs time function was then integrated twice to get the position vs time functions which were  used as the input to the control system. The vertical motion is designed in a way such that the velocity accelerates to a constant $W_0$ before the leading edge of the plate touches the still water surface and then starts to decelerate once the leading edge touches water surface, as shown in Figure \ref{fig:motion}. Since at the moment when the leading edge touches the water surface, the decceleration is nearly zero and starts to increase gradually following a hyperbolic tangent function, the decrease in velocity during this transitional period is small and not easily visible in Figure \ref{fig:motion}. The experiments were performed with four different vertical motions. If we define the Froude number based on the initial impact velocity $W_0$ based on the width of the plate $B$, $Fr=W_0/\sqrt{gB}$, the four experimental condition can be represented by $Fr=0.32,~0.42,~0.53,~0.63$. 

\subsection{Methodologies and setup of Viscous Computations}

In this section we will outline a series of two-dimensional viscous flow computations of a plate impacting the water surface. All the flow parameters and kinematics are the same as the ones in the experiments above.  As we will demonstrate in the following sections, the two-dimensional computations reproduce the experimental observations in a qualitative sense.  To expand the parametric range considered in the experiments, we have also conducted a series of numerical experiments with flexible plates.  These computations revealed some interesting aspects of the effects of the flexibility of the plate on the spray formation, which also serves as a guide for setting up future experiments with flexible plates.

The Navier-Stokes equations for viscous incompressible flow govern the
dynamics of the air and water phase:
\begin{align}
\label{eq:ns}
\frac{D {\bf{u}}}{D t} & =  -\frac{1}{\rho (\phi)} \nabla p + \frac{1}{\rho (\phi)} \nabla \cdot (\mu(\phi) \nabla {\bf u}) + {\bf f}, \\
\nabla \cdot {\bf{u}} & = 0,
\end{align}
where ${\bf u}$ and $p$ are the velocity vector and pressure, respectively. $t$ is time, and ${\bf f}$ is the body force density.  $\rho(\phi)$ and $\mu(\phi)$ are the density and dynamic viscosity of the fluid, respectively. $\phi$ is a level set function to capture the interface between air and water.  We should note that the incompressibility assumption for the air phase is fairly accurate for large impact angles, as the ones considered here.  As the impact angle gets very small, however, it may induce errors.
%
%
In the general case the flexible plate is modeled as a linear
Euler-Bernoulli beam,
\begin{equation}
  \rho_s A \frac{\partial^2 w}{\partial t^2} + E I \frac{\partial^4 w}{\partial x^4} = q + \rho_s A a,
\end{equation}
where $\rho_s$ is the density of the solid body, $w$ is the
displacement, $A$ is the cross section area of the plate, $E$ is the
Young's modulus, $I$ is the second moment of the area of the cross
section along the neutral line, $q$ is the distributed load, $a$ is
the body force density.  

The above set of equations is solved as a coupled system using the
high-fidelity, Navier-Stokes solver for multiphase incompressible
flows, SPLASH. The solver employs level set techniques to sharply
define the interface between different phases
\citep{Qin:2015}. Adaptive Mesh Refinement (AMR) is used to distribute
computational resources in a cost/efficient manner and the mesh is
selectively refined around the interface, which is evolving in time
\citep{Vanella:2010}. A fractional step method is used to solve the
momentum and continuity equations, which results in a variable
coefficient Poisson pressure equation. Proper jump conditions are
applied to the Poisson pressure equation to accurately capture the
jump in pressure that results from surface tension between different
phases. The solver is designed to handle multiphase flows with large
density ratios, such as air and water whose density ratio is
1:1000. The rigid or flexible plate is introduced through the
immersed-boundary formulation proposed by \cite{Vanella:2009}. The
spatial derivatives in the Euler-Bernoulli beam equation are treated
implicitly, and discretized by the 4th-order center difference scheme.
Both fluid and structure are advanced in time simultaneously using a
strong coupling scheme \citep[see][for details]{Yang:2008fv}.




The two-dimensional numerical simulations below mimic the experiments,
where the flat plate impacts vertically the free surface with a
deadrise angle of $10^\circ$. The impact velocity used in the
experiments is prescribed.  The computational domain is $ [-1.0 B,
5.4B] \times [-2.52B, 3.88B] $ in the $y$ and $z$ directions, where
$B$ is the chord length of the flat plate.  The non-slip boundary
condition is used on all the boundaries. The flow is initially static
with the free surface at $z = 0$. The minimum spacial grid is $dh =
B/160$.
The numerical simulations use the same kinematics and physical
parameters as these in the experiments, except that the Reynolds
number is set to be $Re=U_{ref}L_{ref}/\nu_{w} = 10,000$ to keep the
2D simulation stable and assuming that the Froude number dominates the
flow.  The reference length, velocity, time and density are the plate
width ($B$) , $\sqrt{gB}$, $\sqrt{B/g}$, and water density ($\rho_w$),
respectively.

The flexural rigidity of the beam is normalized by $(EI)^* =
EI/(\rho_w U_{ref}^2 L^4)$. This work uses two flexural rigidities,
$(EI)^* = 0.08$ and $(EI)^*=0.64$, which corresponds to the steel
plate with thickness $1 mm$ and $2 mm$, respectively. The density of
the flexible beam is $\rho_s/\rho_w = 8.00$.

\subsection{ILES Methodology and Setup}

\begin{figure}
\begin{center}
\includegraphics[width=0.9\linewidth]{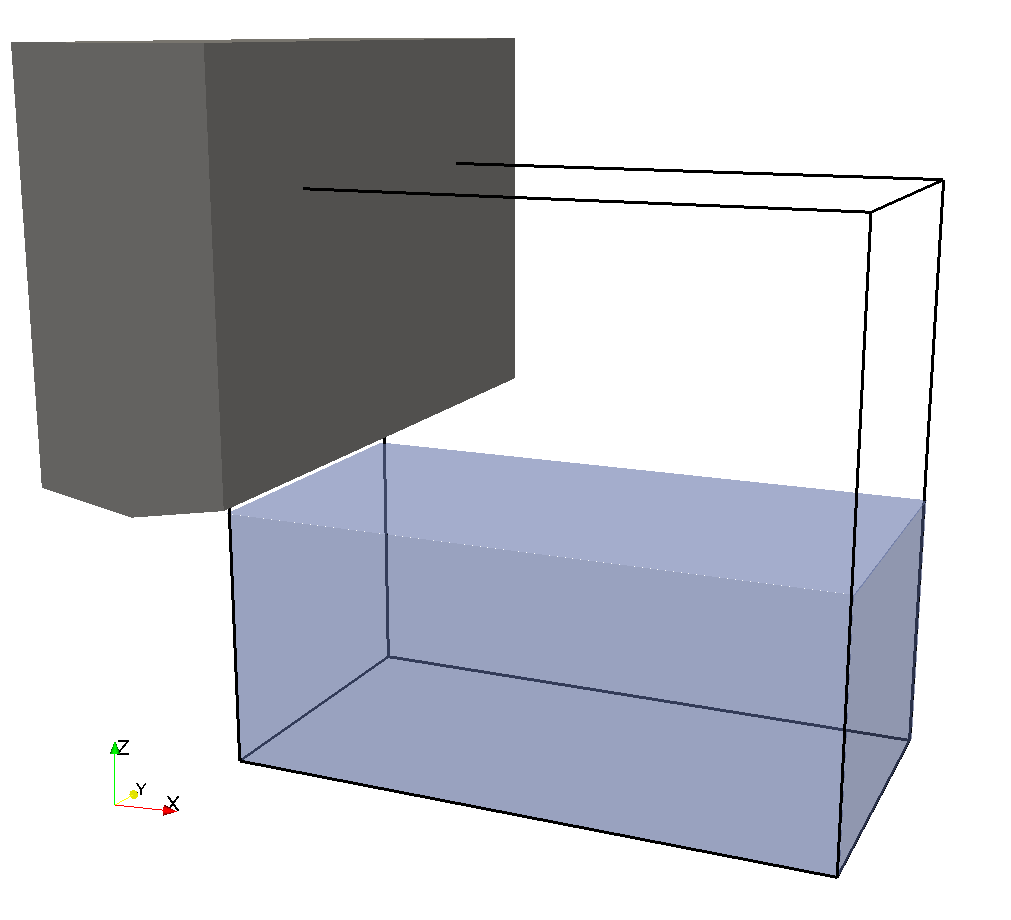}   \\
\end{center}
\vspace*{-.25in}
\caption{NFA simulation domain. 2.44 m by 0.96m by 1.219 m. \label{fig:NFAdomain}}
\end{figure}

 \begin{figure*}[!htb]
\begin{center}
\begin{tabular}{cc}
(a)&(b)\\
\includegraphics[trim={0.5in 0in 0in 0.2in},clip = true,width=0.49\linewidth]{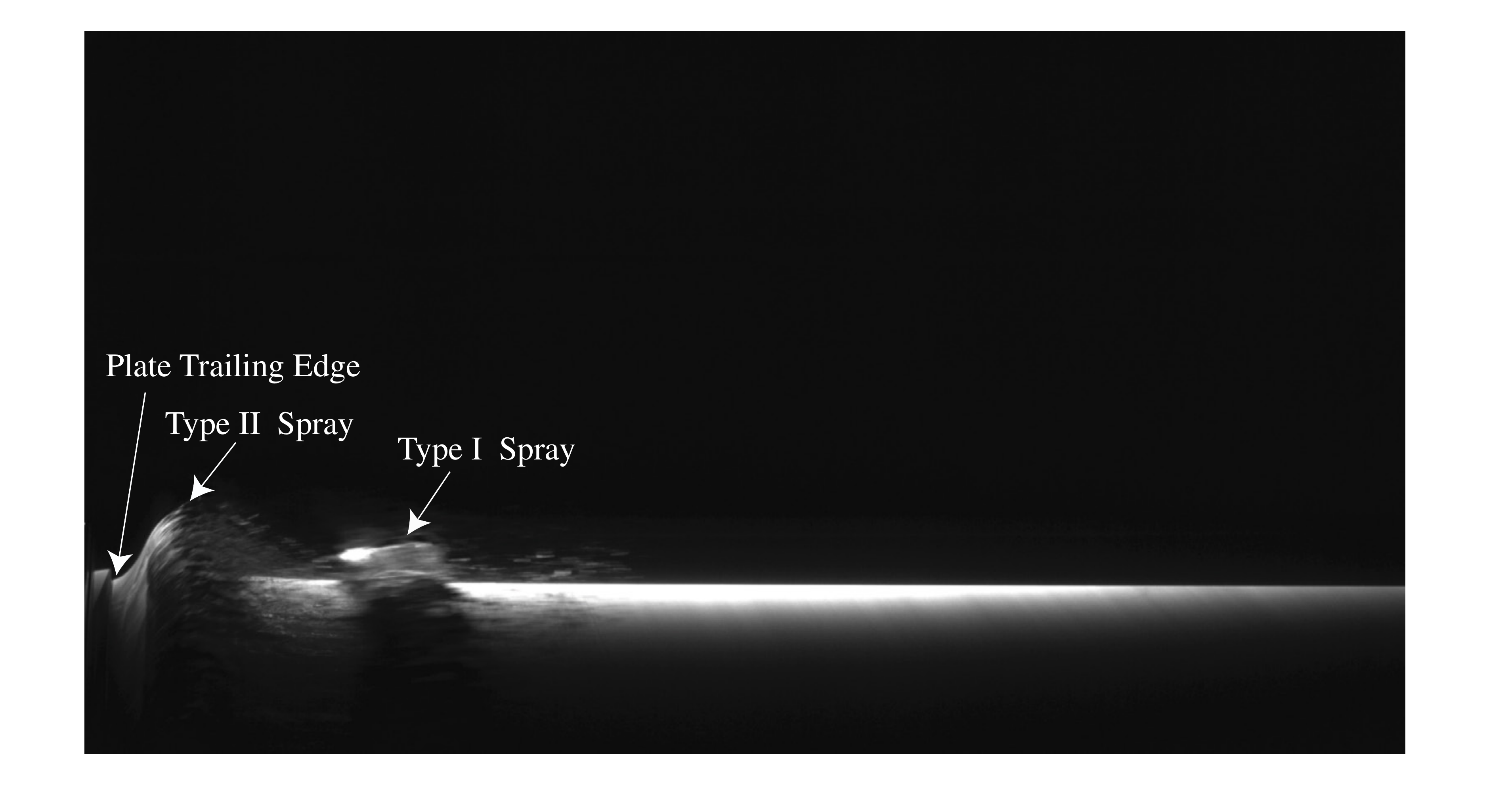}& \includegraphics[trim={0.5in 0in 0in 0.2in},clip = true,width=0.49\linewidth]{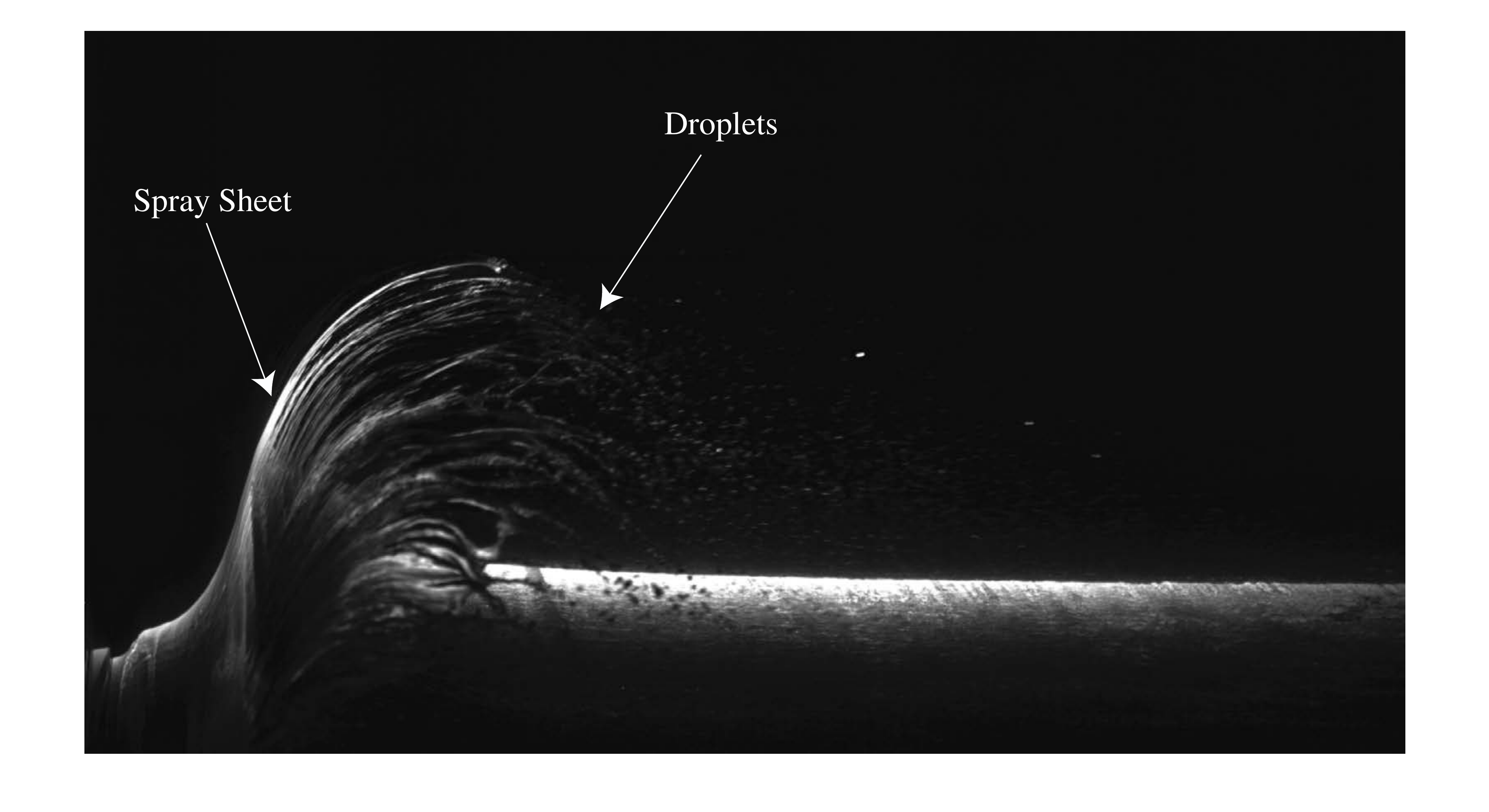}\\
 (c)&(d)\\
 \includegraphics[trim={0.5in 0in 0in 0.2in},clip = true,width=0.49\linewidth]{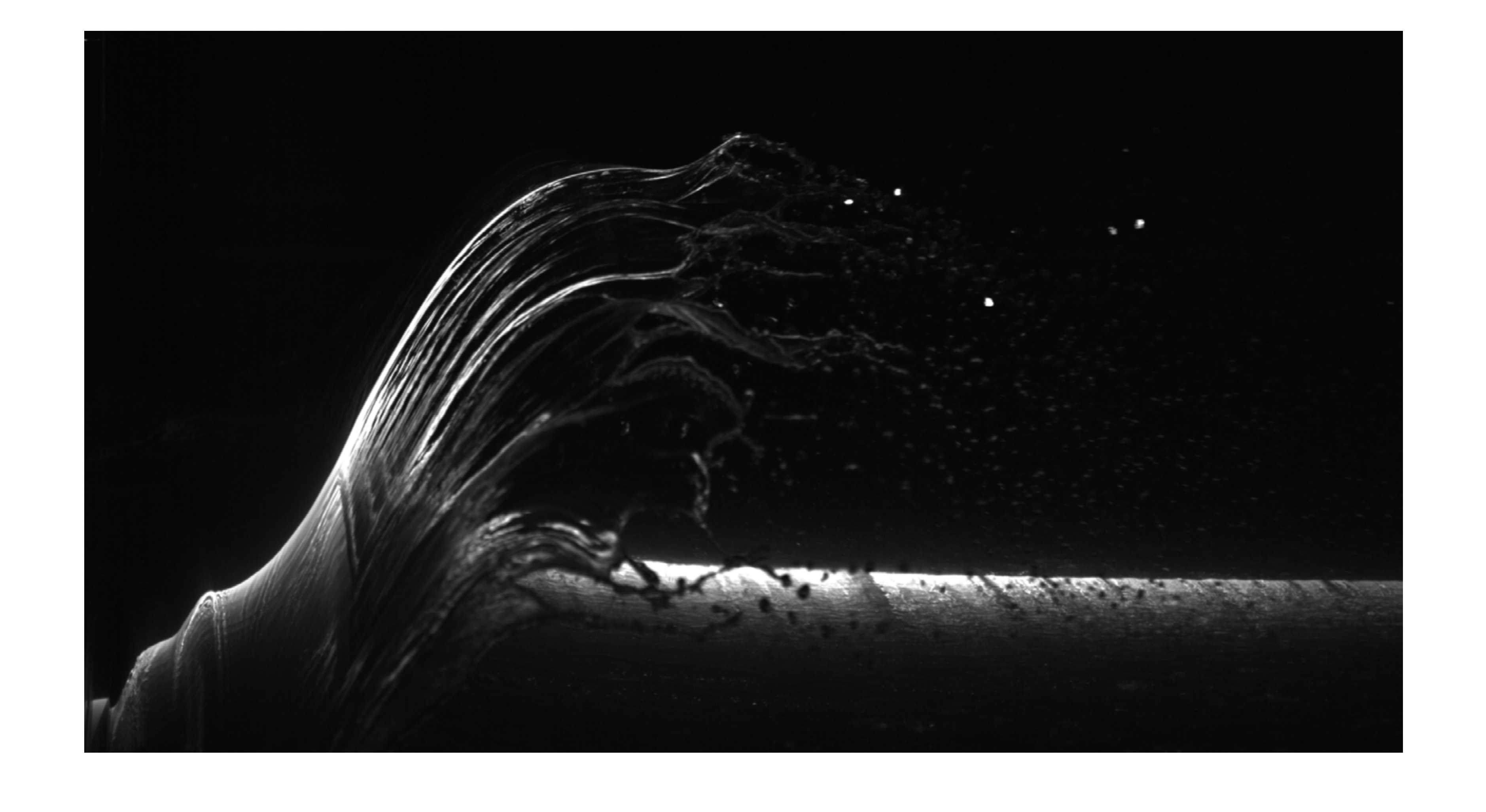}& \includegraphics[trim={0.5in 0in 0in 0.2in},clip = true,width=0.49\linewidth]{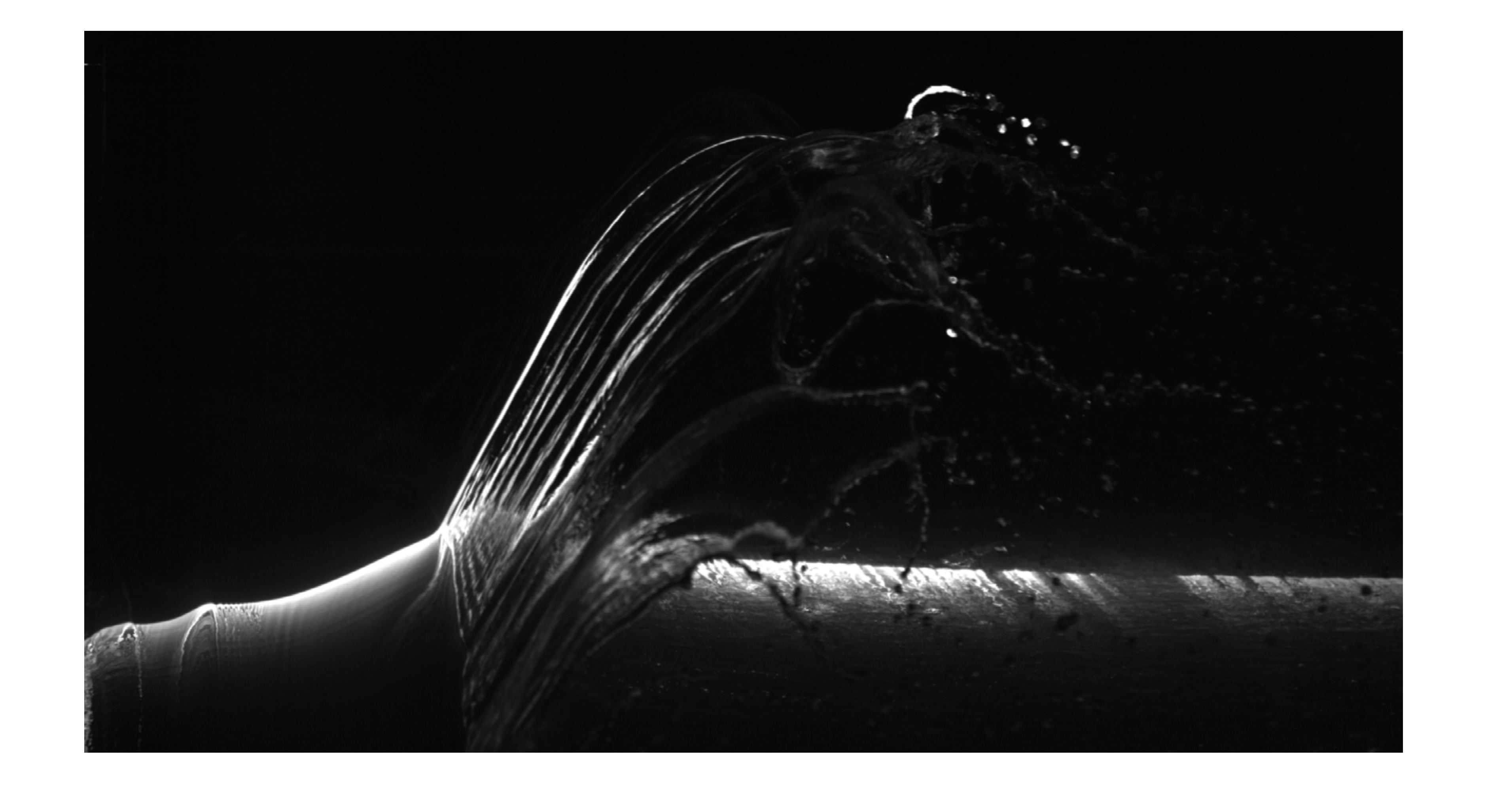}\\
  (e)&(f)\\
 \includegraphics[trim={0.5in 0in 0in 0.2in},clip = true,width=0.49\linewidth]{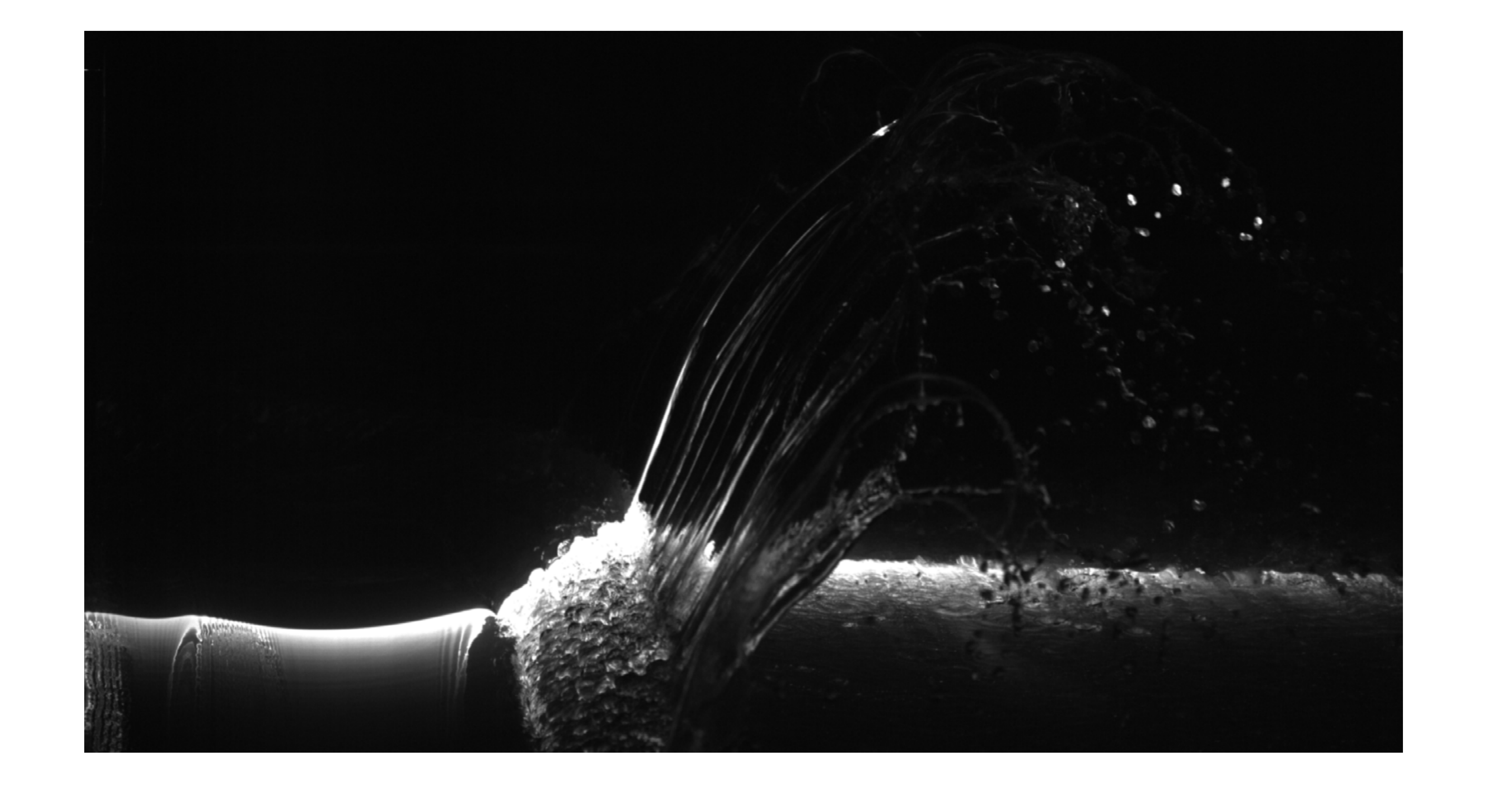}& \includegraphics[trim={0.5in 0in 0in 0.2in},clip = true,width=0.49\linewidth]{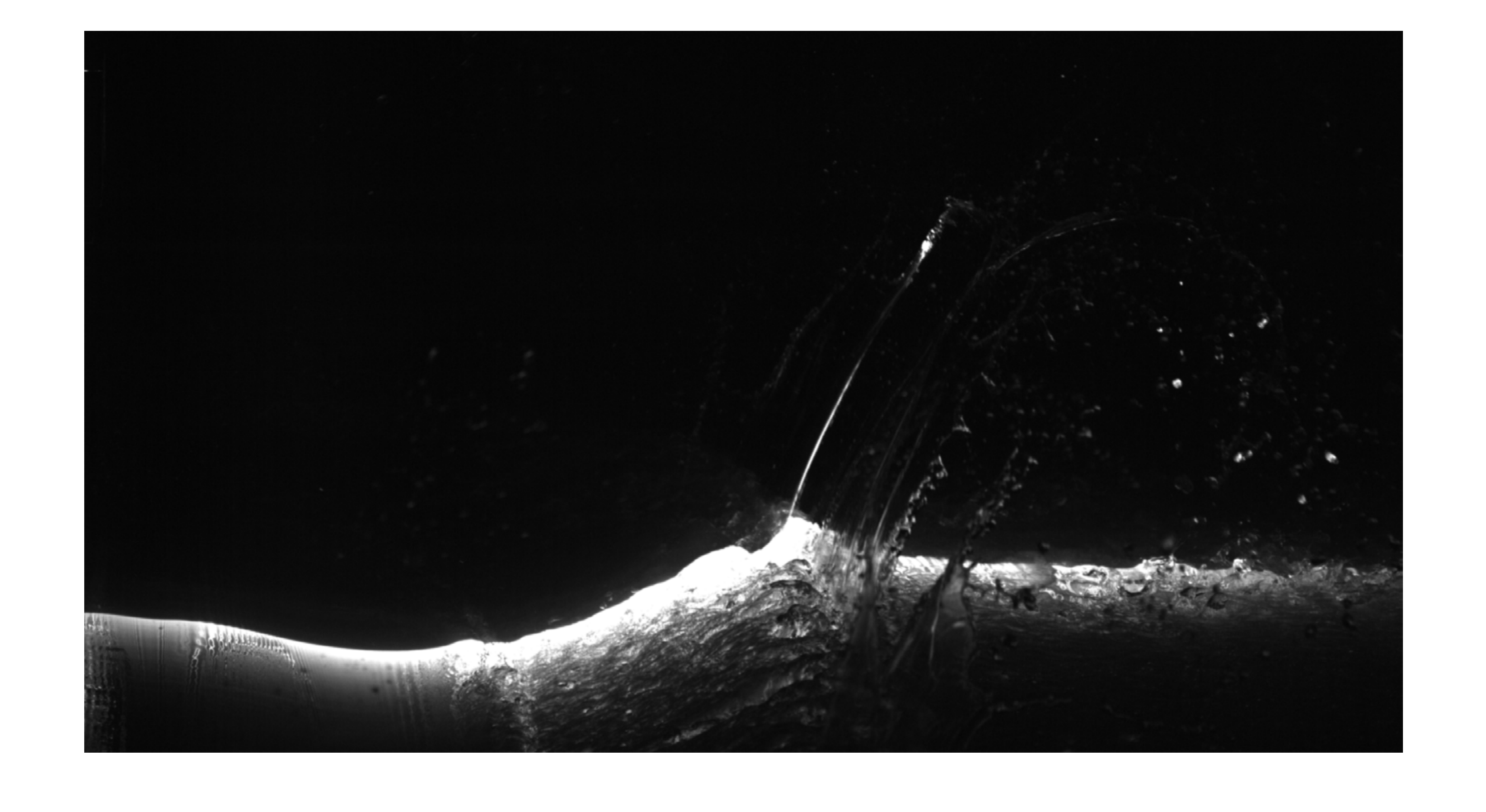}\\
\end{tabular}
\end{center}
\vspace*{-0.25in}
  \caption{Sequence of LIF images of the spray formation for $Fr=0.42$. The field of view is 106~cm$\times$68~cm. The time interval between successive images is 0.1~s. In the LIF images, the sharp boundary between the dark region and bright region represents the air-water interface. The trailing edge of the plate is located at the left side of the images. Type I spray, which consists of a cloud of droplets and ligaments, is shown in (a). Type II spray is a spray sheet originated from the spray root. Evolution of Type II spray is shown from (a) to (f). \label{fig:spray_pic}}
\end{figure*}

NFA provides a numerical capability to accurately model the breaking of waves in the ambient ocean and around ships, including the formation of spray, entrainment of bubbles, and plunging and spilling breaking waves (for example see: \cite{Dommermuth:2006}, \cite{OShea:2008}, and \cite{Dommermuth:2014}).  NFA uses a Cartesian-grid method with the body treated using an immersed boundary method with a surface panelization of a geometry as input. The air-water interface is modeled with the volume of fluid technique and is capable of capturing wave overturning and free-surface turbulence.

NFA is uniquely suited to address problems pertaining to spray and spray sheet formation. Computational Fluid Dynamics (CFD) codes often have inadequate free-surface resolution that can be problematic when capturing small scale features. The violent and small-scale nature of the flow also requires a non-diffusive interface capturing method. The Cartesian grid method utilized by NFA lends itself to simple and extremely effective parallelization. Combined with the Navy's expanding suite of high performance computing resources this allows for large problems to be solved over many thousands of cores, thus reducing turnaround time. The explicit time stepping scheme used in NFA also facilitates parallelization while the ILES formulation ensures stability for complex multiphase flows. The upwinding in NFA's advective terms has the effect of stabilizing the explicit time stepping and allows for fewer matrix solves per time step. Using an immersed boundary, cut cell method for our treatment of the body allows for rapid integration of complex and novel hull designs.
 
NFA uses a density weighted velocity smoothing scheme to control the breakup of the free surface. The free-surface boundary layer is not resolved in volume of fluid (VOF) simulations at high Reynolds numbers with large density jumps such as air and water. Under these circumstances, the tangential velocity is discontinuous across the free-surface interface and the normal component is continuous. As a result, unphysical tearing of the free surface tends to occur. Mass-weighted filtering can be used to alleviate this problem by forcing the air velocity slightly above the interface to be driven by the water velocity slightly below the interface in a physical manner. Details of the algorithm can be found in \cite{nfa1}.
 
We simulated the vertically moving plate slamming experiment in NFA using a solid rectangular prism and a domain size that is consistent with the tank width and depth, 2.44 m and 0.96 m respectively. The length of the domain along the longer side of the plate is equal to the plate length of 1.219 m. A representation of the domain and geometry are shown in Figure  \ref{fig:NFAdomain}. The quasi 2D nature of the flow allowed us to cluster points toward the center of the plate. The images presented in this paper are from that region of high density.
  
The focus of the NFA section for this paper is on the case with a plate impact velocity of 24 inches per second. In order to adequately represent the thin spray sheet and spray formation high grid resolution is required. To achieve this a grid spacing of 1.2 mm is used near the plate and in the area of the spray sheet propagation. A minimum time step of 5.0E-05 seconds was necessary due to the high velocities and small grid spacing. The CFL number is set to 0.5. The number of grid points used along the width, length and depth of the domain are 1088, 512, and 768 respectively. The total simulation used 428 million grid cells and was run on the Navy's IBM iDataplex Haise using 1632 CPUs. The simulation had a run time of 12 hours. The plate started at rest and was accelerated to a top speed of 24 inches per second based on experimentally measured position data. Flow field velocities are initialized at zero.

\section{Results and Discussion}
\subsection{Experiments}

\begin{figure}
\begin{center}
\vspace*{-0.5in}
 \includegraphics[trim={1.2in 0in 0in 0in},clip=true,width=1.08\linewidth]{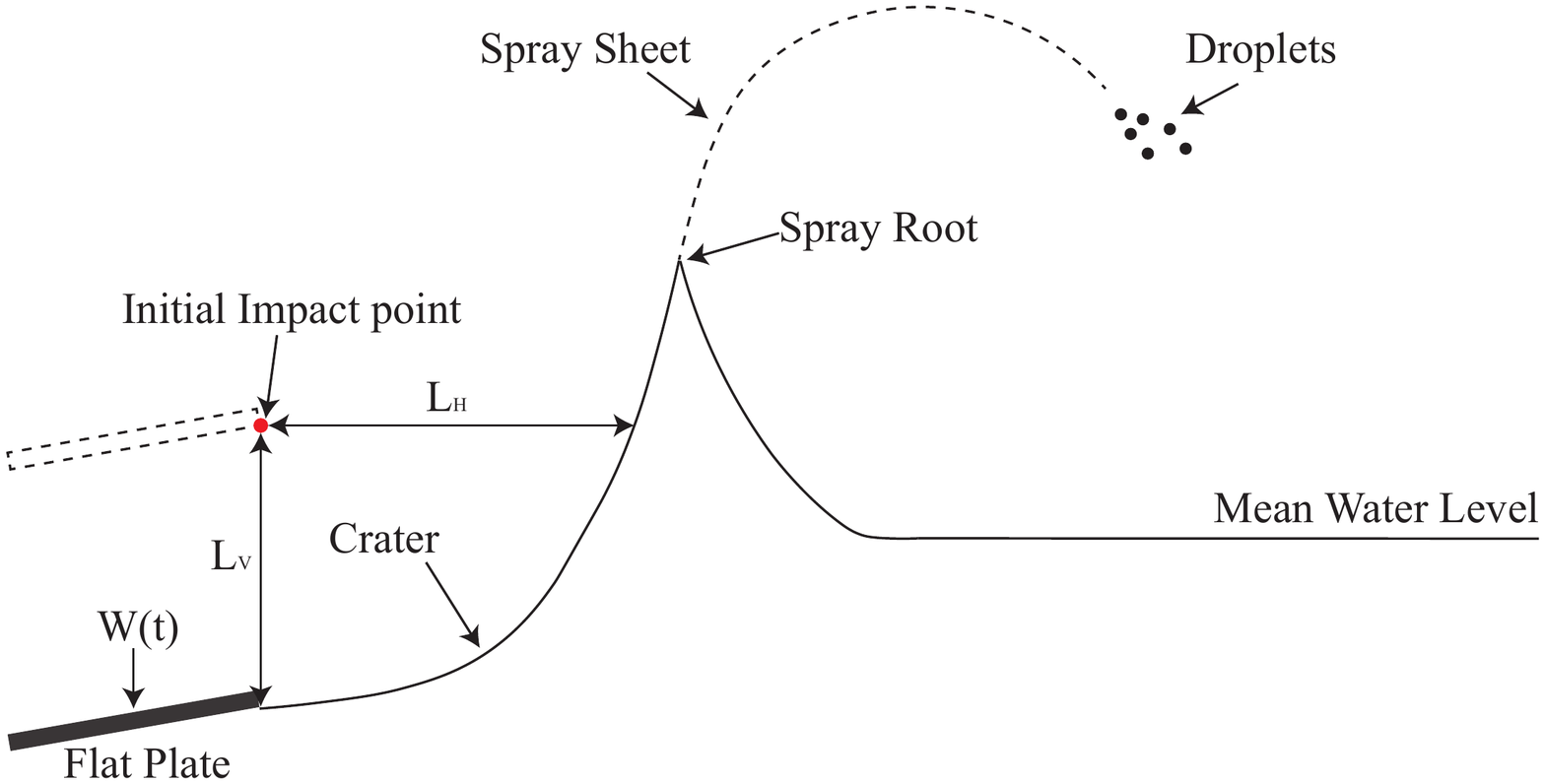}
\end{center}
\vspace*{-0.85in}
  \caption{ A schematic diagram showing various features of the free surface shape during plate impact.  The initial impact point is the point where the trailing edge of the plate first touches the local water level. $L_H$ and $L_V$ are the horizontal and vertical distance from the initial impact point to the surface of crater, respectively.\label{fig:SpraySchem}}
\end{figure}

\begin{figure*}[!htb]
\begin{center}
\begin{tabular}{cccc}
(a)&(b)&(c)&(d)\\
 \includegraphics[trim={0in 0in 0in 0.1in},clip = true,width=0.245\linewidth]{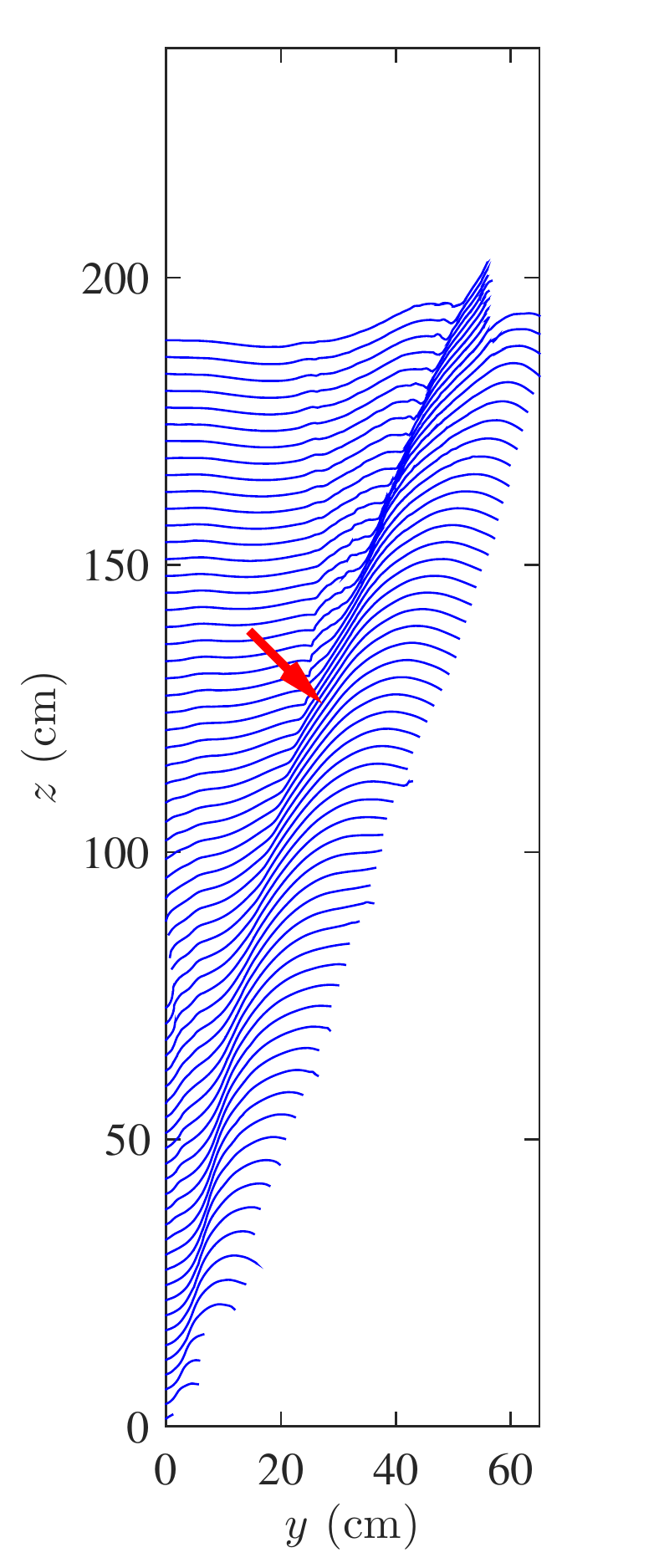}&\includegraphics[trim={0in 0in 0in 0.1in},clip = true,width=0.245\linewidth]{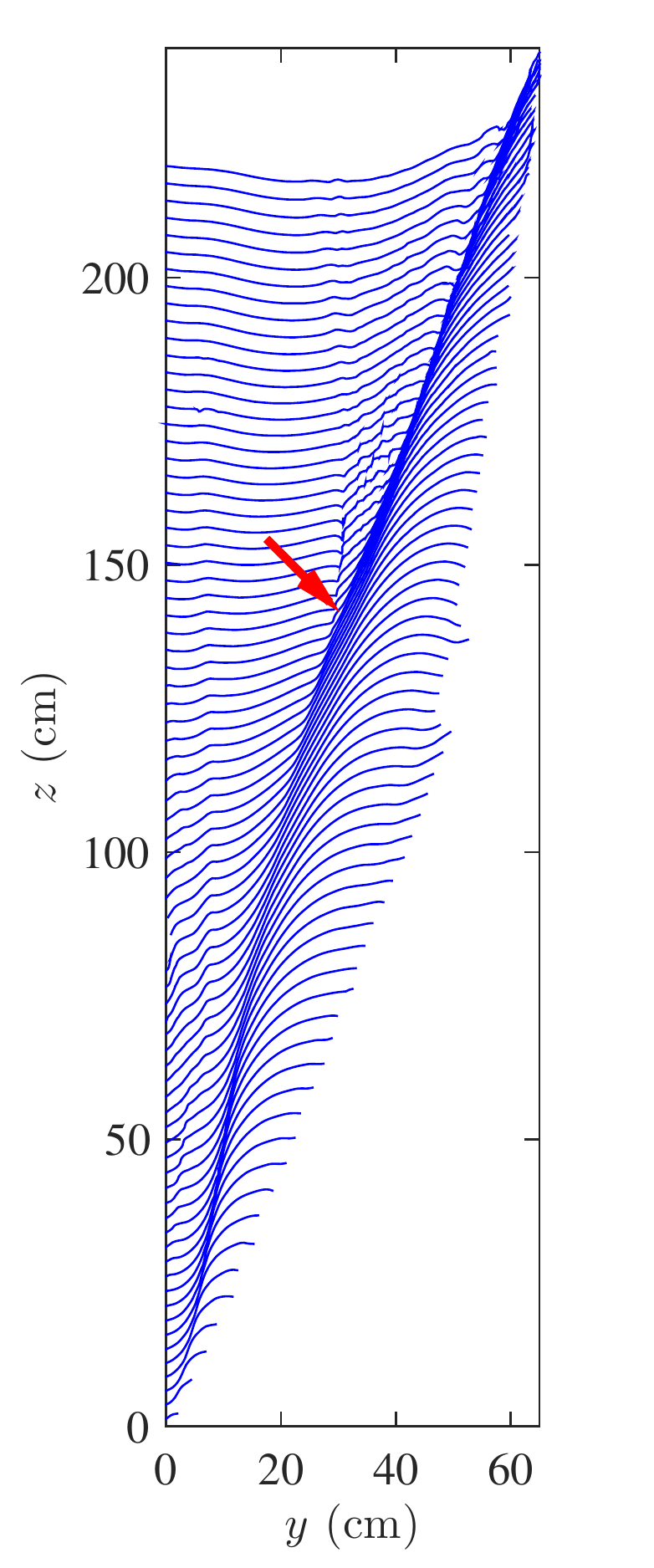}&\includegraphics[trim={0in 0in 0in 0.1in},clip = true,width=0.245\linewidth]{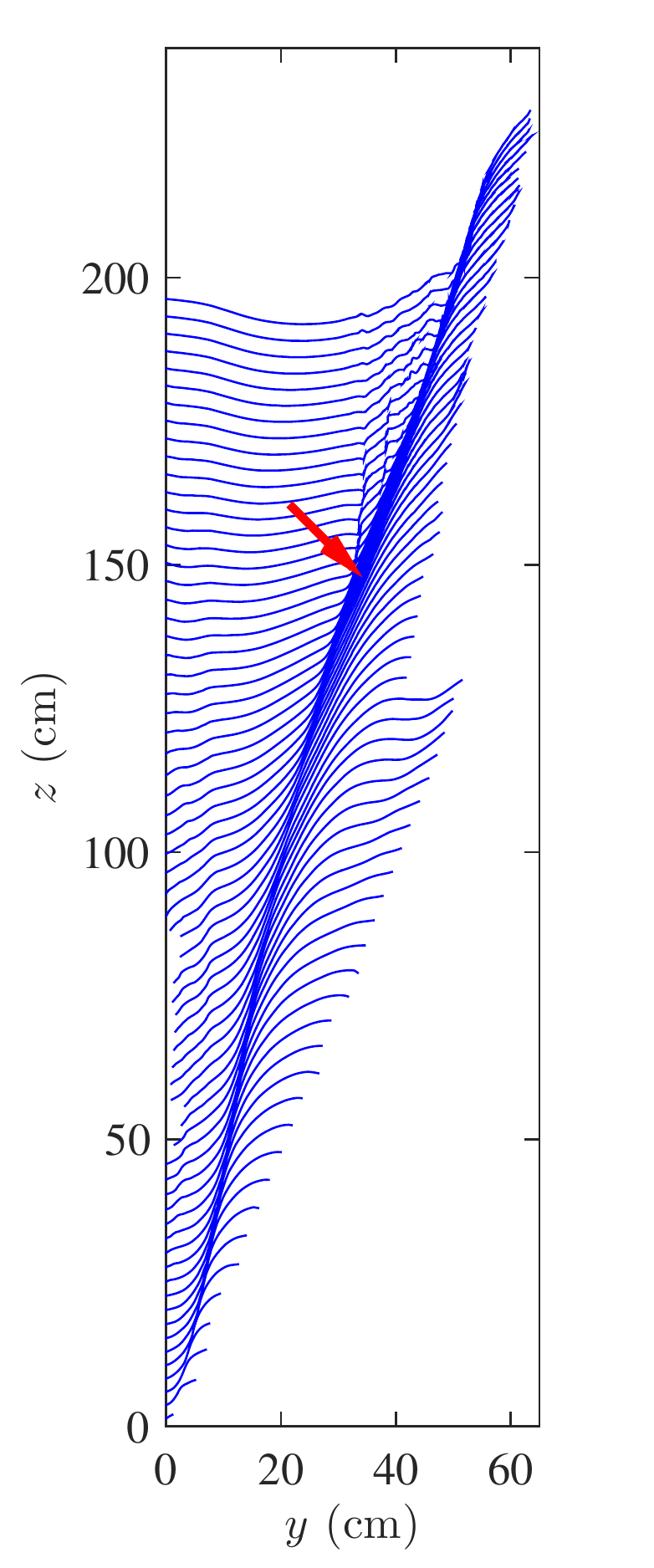}&\includegraphics[trim={0in 0in 0in 0.1in},clip = true,width=0.245\linewidth]{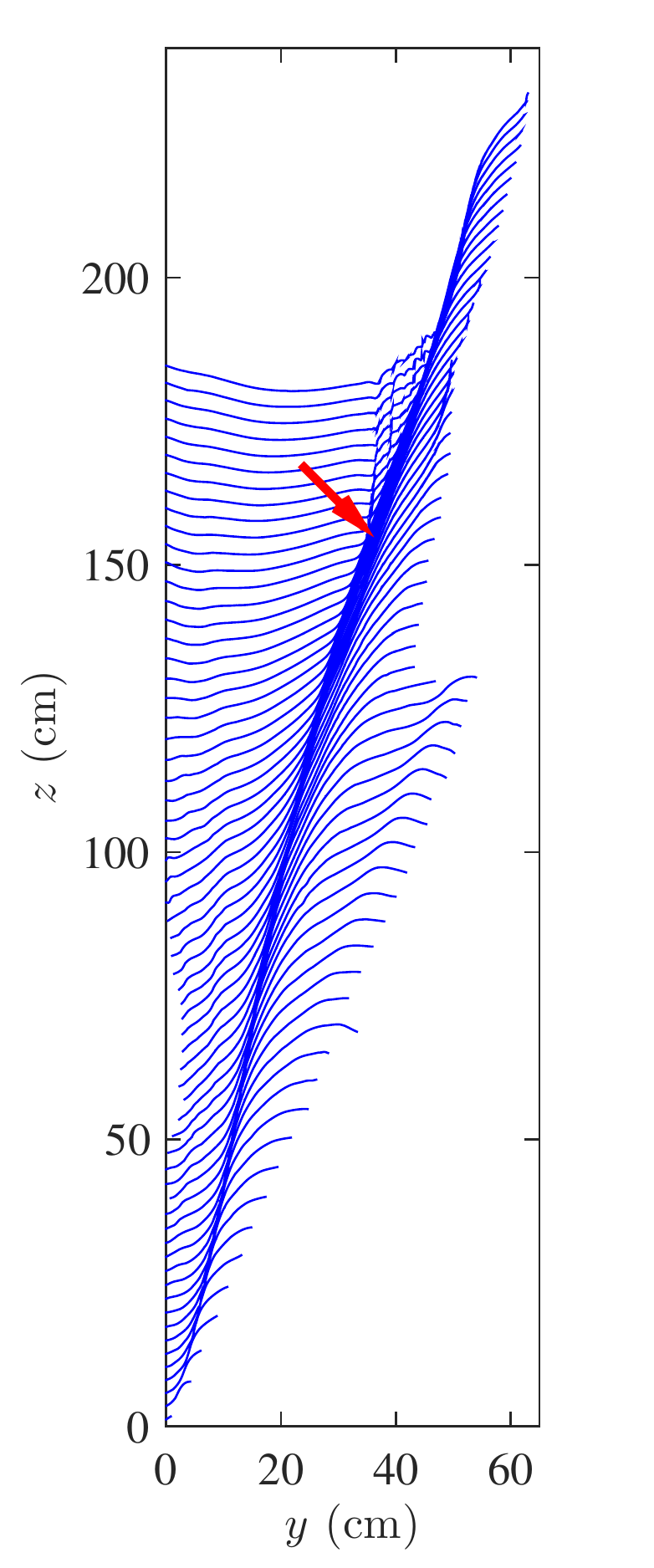}
\end{tabular}
\end{center}
\vspace*{-0.3in}
  \caption{Surface profiles at different impact velocities. Each subsequent profile is shifted upward by 1.5~cm from the previous profile. The trailing edge of the plate is located at $y=0$. The profile is plotted from the trailing edge of the plate to a point where the spray sheet is still well defined in the LIF images. From (a) to (d), the profiles are measured from $Fr=0.32,~0.42,~0.53,~0.63$, respectively. The red arrow in each figure points to the spray root point of the frame when the turbulent splash zone starts to form. \label{fig:profiles}}
\end{figure*}

A sequence of images from the high-speed LIF movie for $Fr=0.42$ are shown in Figure \ref{fig:spray_pic}. The field of view is roughly 106~cm$\times$68~cm. The time interval between each image is 0.1~s. The trailing (high) edge of the plate is at the left of the image. Figure \ref{fig:spray_pic}(a) shows a moment shortly after the trailing edge of the plate touches the local water surface. A cloud of droplets and ligaments moves with high horizontal velocity across the tank. This spray is defined as Type I spray, which is originated from the leading (low) edge of the plate when it first impacts the water surface. Shortly after the Type I spray is generated, there is a rise of the local water level near the trailing edge of the plate, due to the water entry effect under the plate. Then, as the plate continues to move down, the trailing edge of the plate first touches the local water surface and produces the Type II spray as shown in Figure \ref{fig:spray_pic}(a). A schematic diagram showing various features of the water surface profile for the Type II spay is given in Figure \ref{fig:SpraySchem}.   The Type II spray consists of a thin continuous spray sheet of water that originates from a spray root point.  The overall scale of the spray sheet at first grows in time. At the far end of the spray sheet, the sheet becomes unstable and breaks up into droplets, as shown in Figure \ref{fig:spray_pic}(b)(c)(d). Eventually the spray sheet falls down under the effect of gravity and collapses on the free surface, forming a free surface turbulent splash zone, as shown in Figure \ref{fig:spray_pic}(e)(f). 

The surface profiles extracted from the high-speed LIF movies for four different Froude numbers are shown in Figure \ref{fig:profiles}. The profiles are measured from the trailing edge of the plate to a point on the profile where the surface becomes ill defined due to breakup of the spray. In the plot, each profile is shifted upward by 1.5~cm from the pervious profile in the physical plane for clarity of the presentation.   The first profile is the surface profile just after the trailing  edge of the plate impacts on the local water surface and the  time interval between successive profiles is 7.5~ms. In Figure \ref{fig:profiles}, the profiles around the spray root  form a region with high line density because of the high curvature of this portion on each profile. This region with high line density qualitatively shows the evolution of the horizontal position of the spray root. The trajectory of the spray root  seems to be similar qualitatively for the four values of $Fr$. The free surface turbulent splash zone (seen to the left of the spray root at about 150~mm on the vertical axis) starts progressively  earlier in time with decreasing $Fr$, because the higher $Fr$ cases generate larger spray, which takes a longer time to collapse.

\begin{figure}[!htb]
\begin{center}
\includegraphics[width=0.95\linewidth]{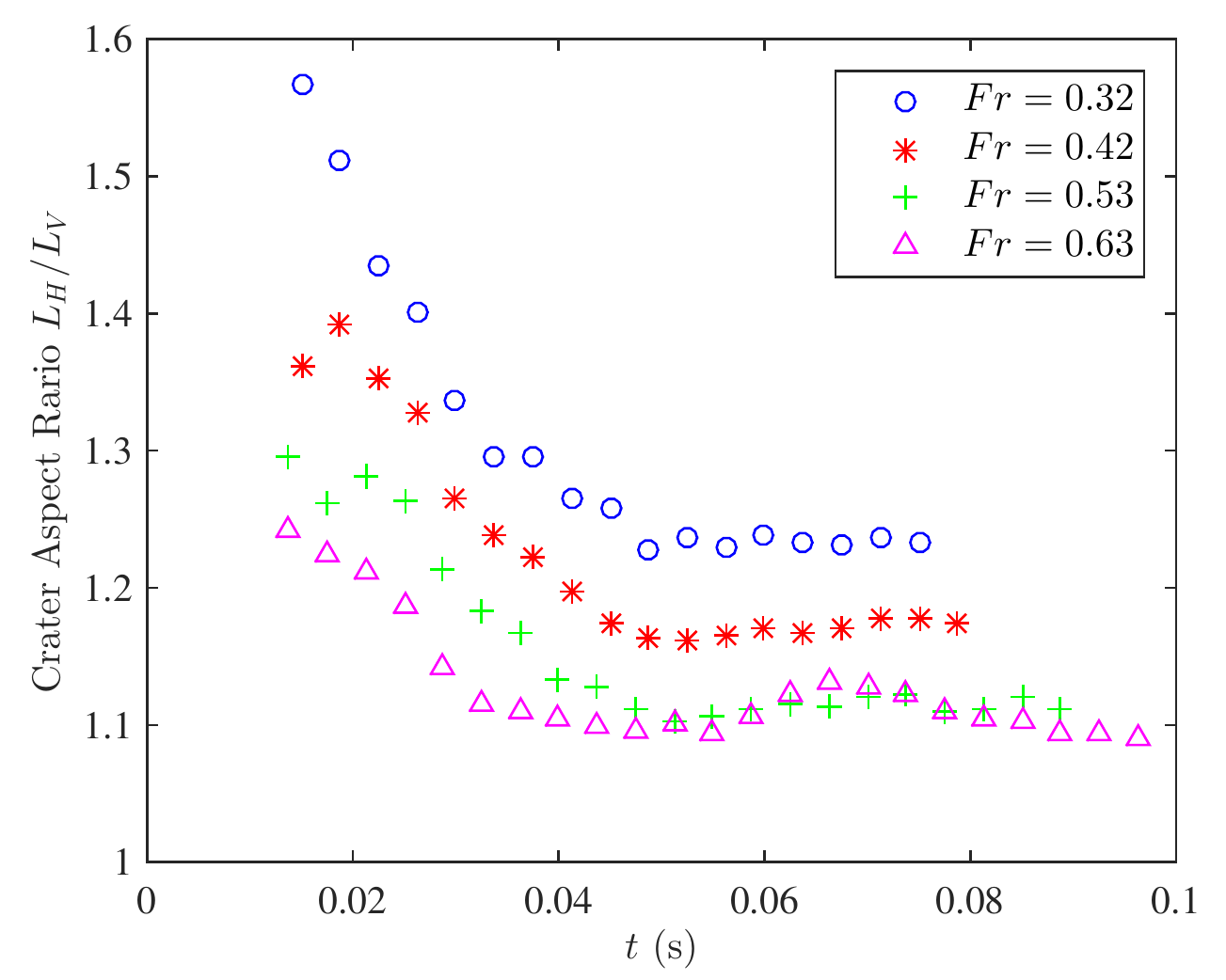}\\
\end{center}
\vspace*{-0.3in}
  \caption{The aspect ratio of the crater ($L_H/L_V$) vs time for the four values of $Fr$.  $t=0$ is the time of the initial impact of the trailing edge.  \label{fig:crater}}
\end{figure}

As depicted in Figure \ref{fig:SpraySchem}, starting from the trailing edge of the plate, a crater is formed and grows as the plate keeps moving downward. One end of the crater is in contact with the trailing edge of the plate while the other end is connected to the spray root point.  Below the spray root point, as the crater grows in size, the shape of the surface is analogous to a propagating wave crest. The intersection of the trailing edge of the plate and the local water surface when they first meet is defined as the initial impact point. If the horizontal distance from the initial impact point to the crater surface is taken as $L_H$ and the vertical distance from the initial impact point to the crater surface is taken as $L_V$, the aspect ratio of the crater can be computed as $L_H/L_V$. Values of the crater  aspect ratio vs time for the four values of $Fr$ are shown in Figure \ref{fig:crater}. For all four $Fr$ cases, the aspect ratio decreases initially and reaches a nearly constant value after some time. The initial aspect ratio decreases monotonically with increasing $Fr$, but, in all cases, 
the aspect ratio is always larger than one. 

\begin{figure}[!htb]
\begin{center}
\begin{tabular}{c}
(a)\\
 \includegraphics[trim={0in 0in 0in 0.2in},clip = true,width=0.95\linewidth]{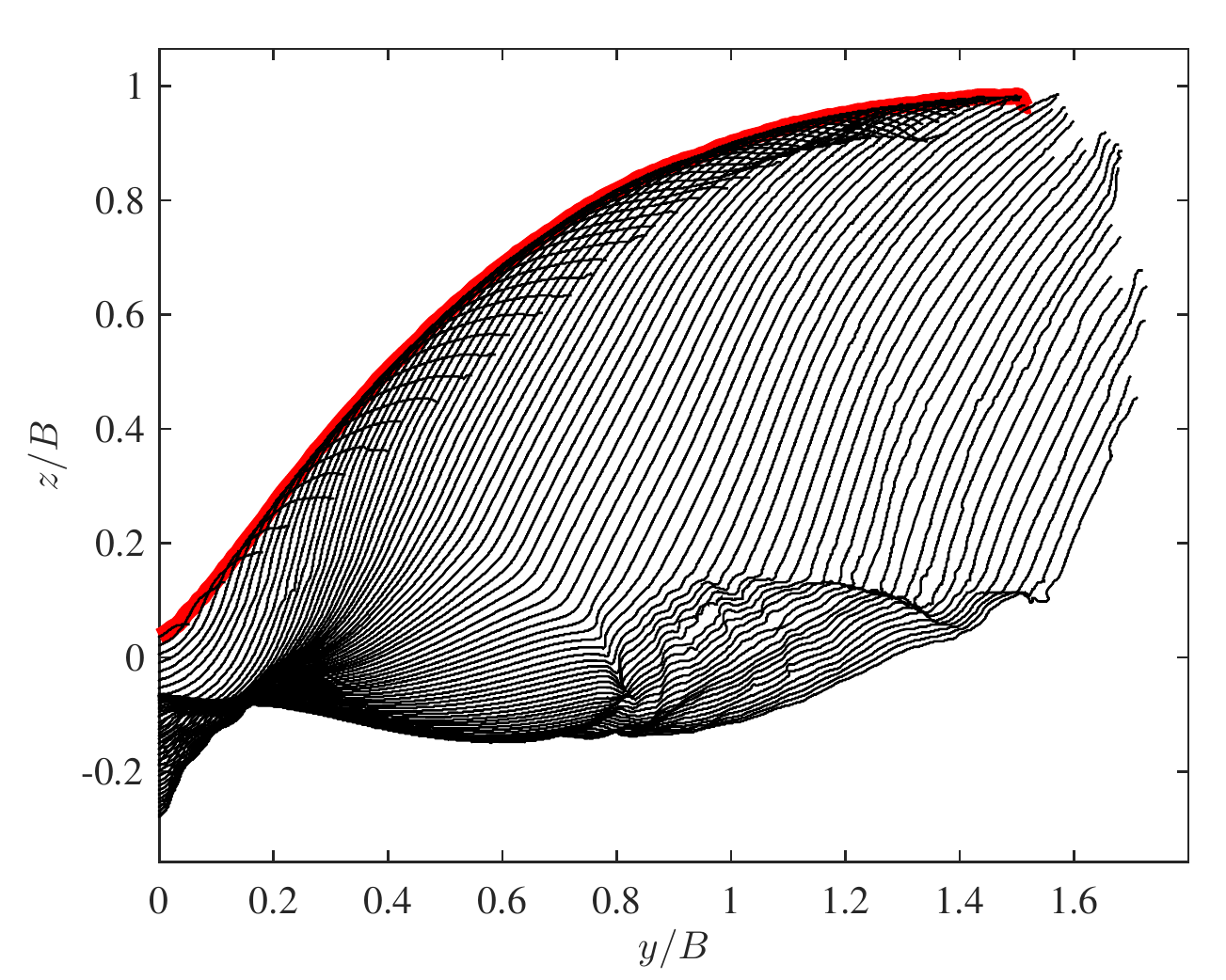}\\
 (b)\\\includegraphics[trim={0in 0in 0in 0.2in},clip = true,width=0.95\linewidth]{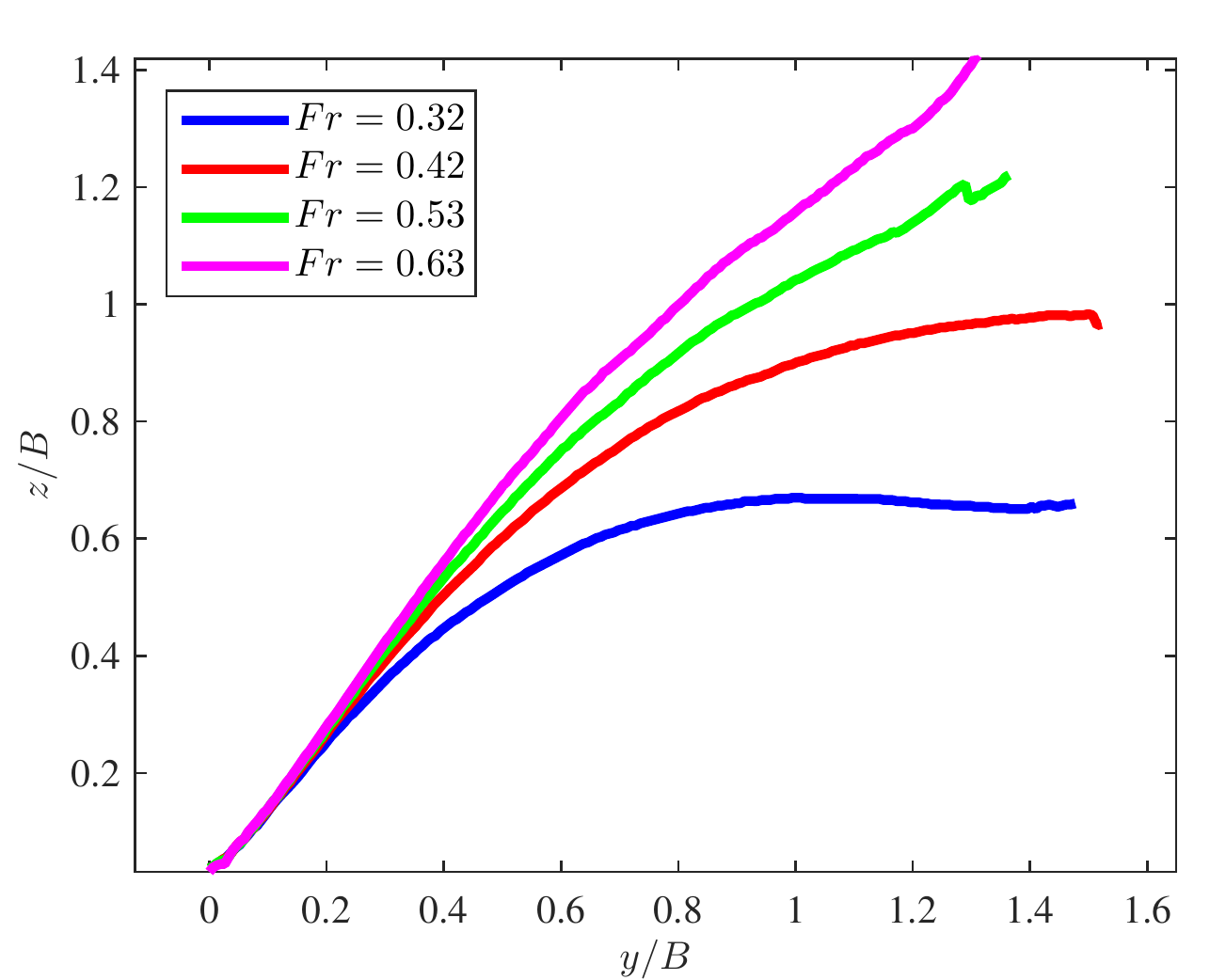}\\
\end{tabular}
\end{center}
\vspace*{-0.25in}
  \caption{Envelopes of the profiles at different Froude number. (a) The envelope (red) of the profiles (black) at $Fr=0.42$. The Envelope is a line that consists of the highest points the spray ever reaches. (b) Envelopes of profiles at $Fr=0.32,~0.42,~0.53,~0.63$.  \label{fig:envelope}}
\end{figure}

An interesting measure of the spray is its envelope, defined here as the curve of the highest positions that the spray ever reaches at each cross-stream, $y$, location.  This envelope is shown as the red line in Figure~\ref{fig:envelope}(a) where the set of profiles for the $Fr=0.42$ case is shown in the laboratory reference frame.
Figure \ref{fig:envelope}(b) shows the evelopes for each of the four values of $Fr$.  
 In the region close to the trailing edge of the plate, the envelopes have nearly the same slope for all  $Fr$. At $y/B=0.3$, the envelopes at different $Fr$ start to deviate from each other while the slope starts to decrease as $y/B$ increases for all four cases. The spray for higher $Fr$ reaches a larger height.

\begin{figure}
\begin{center}
\begin{tabular}{c}
(a)\vspace{-0in}\\
 \includegraphics[trim={0in 0.5in 0.4in 1in},clip = true,width=0.95\linewidth]{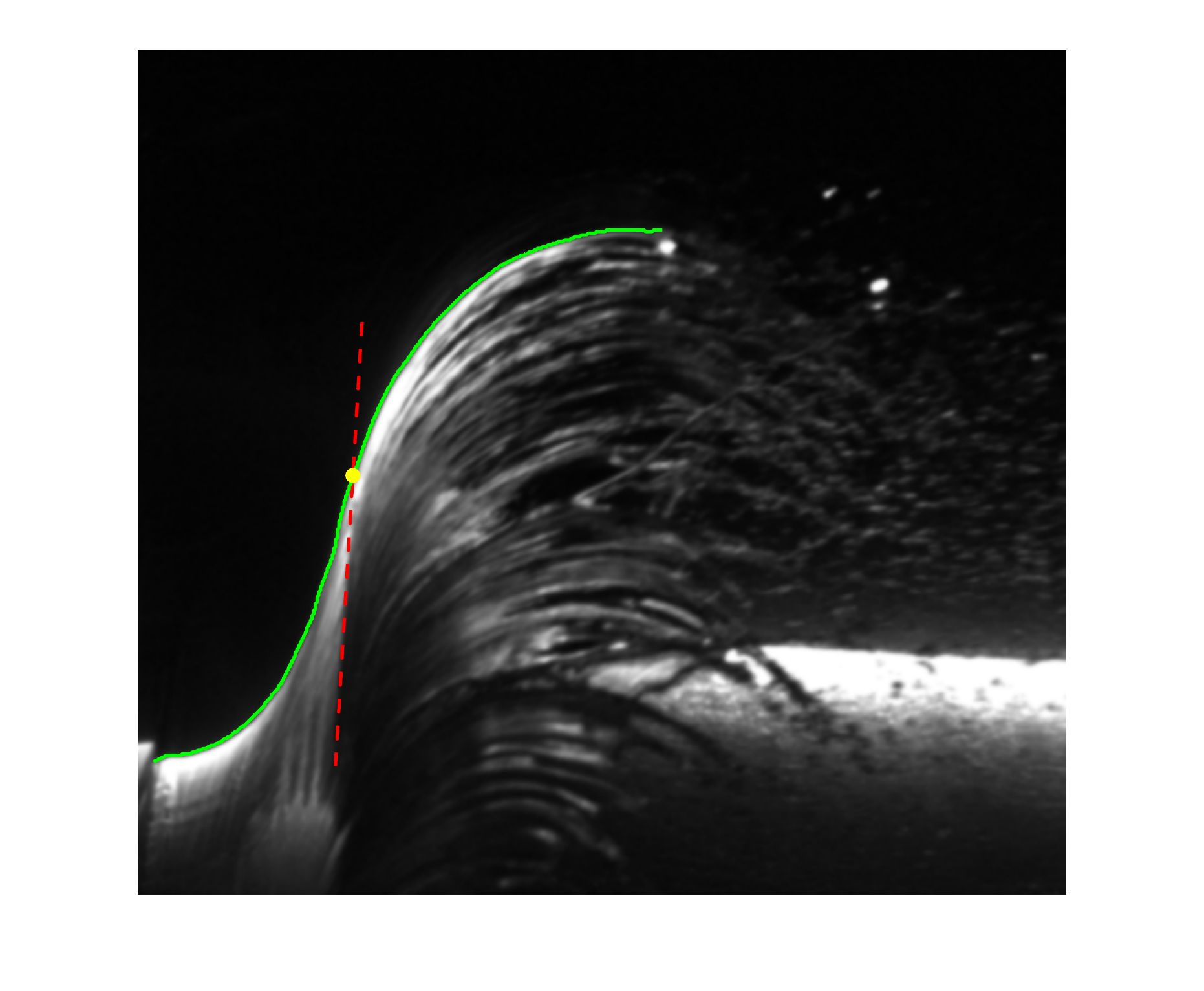}\\
 (b)\\
 \includegraphics[trim={0.1in 0in 0in 0.2in},clip = true,width=0.95\linewidth]{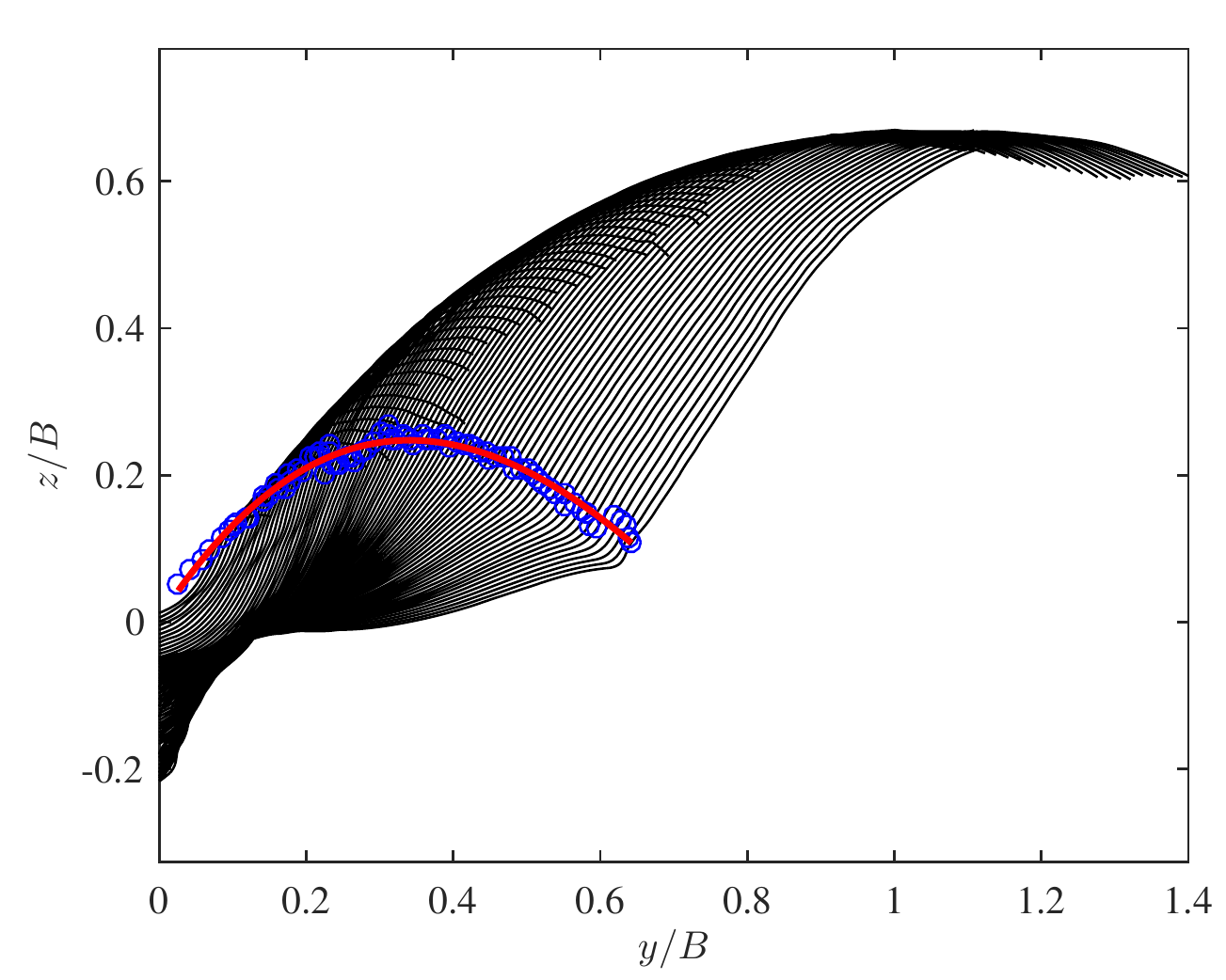}\\
 (c)\\
 \includegraphics[trim={0in 0in 0in 0.2in},clip = true,width=0.95\linewidth]{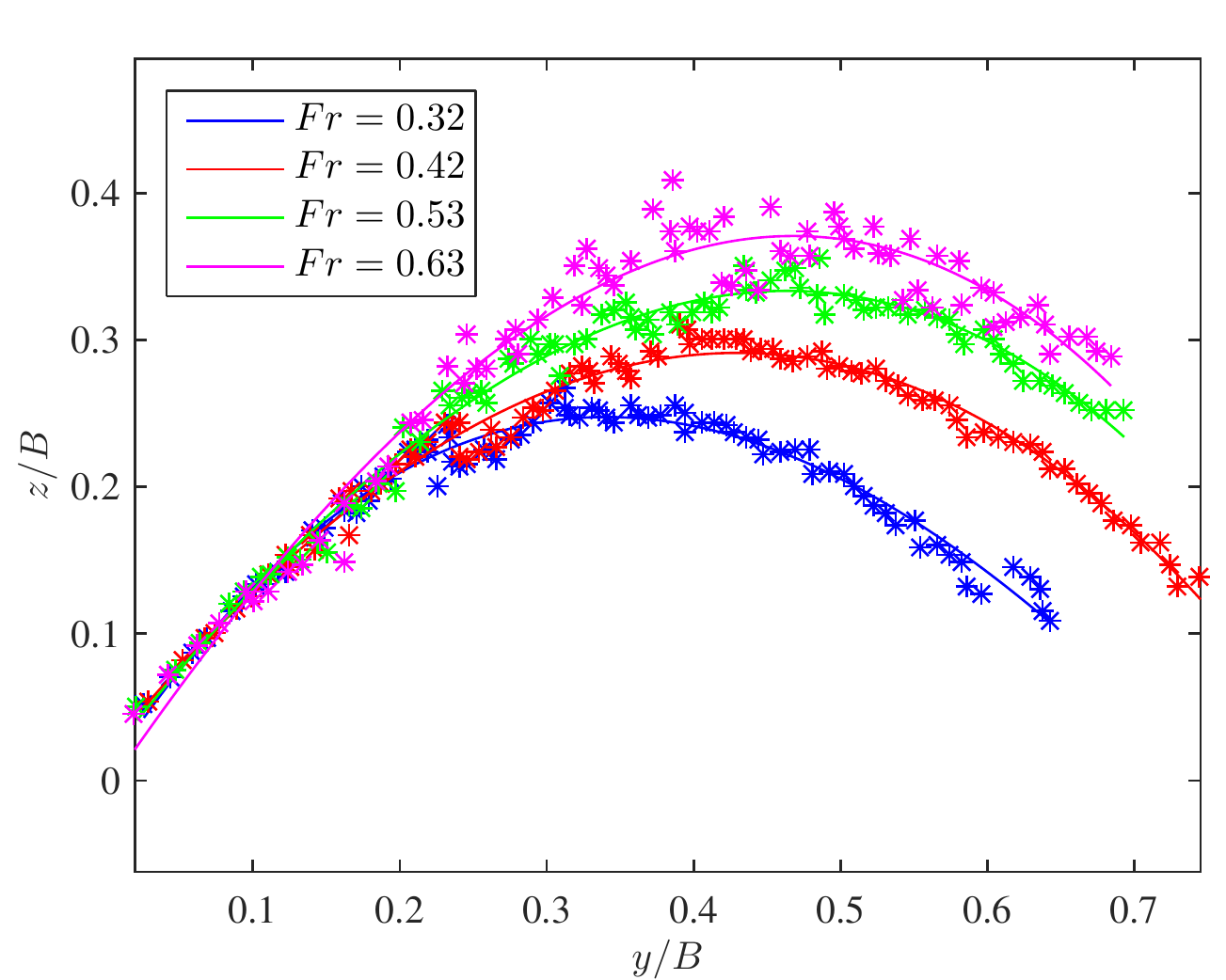}\\
\end{tabular}
\end{center}
\vspace*{-0.3in}
  \caption{Definition of the spray root point. (a) Detection of the spray root point from the LIF image.  The green solid line is the profile of the spray sheet.  The red dashed line is the spray root line while the yellow dot is the spray root point in the plane of the light sheet.  (b) Detected spray root points (blue dots) for  each spray profile, $Fr=0.42$. The red solid line is a 3$^{rd}$ polynomial fit to the detected spray root points. (c) The trajectory of the spray root point for each Froude number. The asterisks are the spray root points detected from the LIF images and the solid lines are the 3$^{rd}$ polynomial fits to these points. \label{fig:sprayroot}}
\end{figure}

The spray root point is defined as the point where the spray sheet originates, as illustrated in Figure \ref{fig:crater}. In three dimensional space, the spray root is a straight line perpendicular to the plane of the light sheet. Because the cameras capture some features of  the water surface shape between the light sheet plane and the camera sensor, the spray root appears as a straight line in the LIF images. For image points below the spray root line, the camera sees through the air-water interface and into the wave below.  For points above the spray root line, the camera sees through the spray sheet into the air underneath it.  These effects result in a high image intensity gradient at the spray root line; therefore, allowing the spray root point to be detected in  the plane of the light sheet.  This detection process is shown in Figure \ref{fig:sprayroot}(a) where the intersection of the surface profile and the extension of the detected spray root line gives the coordinate of the spray root point in the light sheet. The positions of the spray root point are plotted on top of the spray profiles for the $Fr=0.32$ case in Figure \ref{fig:sprayroot}(b), and  the spray root trajectory, as determined by fitting a 3rd order polynomial to the position data, is given as well.  The trajectories of the spray root point for the four values of $Fr$ are shown in Figure \ref{fig:sprayroot}(c). At early times, the trajectories are quite similar.  At later times, roughly when $y/B>0.2$, the trajectories of the spray root diverge from one another; the maximum height and the $y/B$ position of this maximum increase with increasing $Fr$.  

\begin{figure}[!htb]
\begin{center}
\begin{tabular}{c}
(a)\\
 \includegraphics[trim={0in 0in 0in 0.2in},clip = true,width=0.95\linewidth]{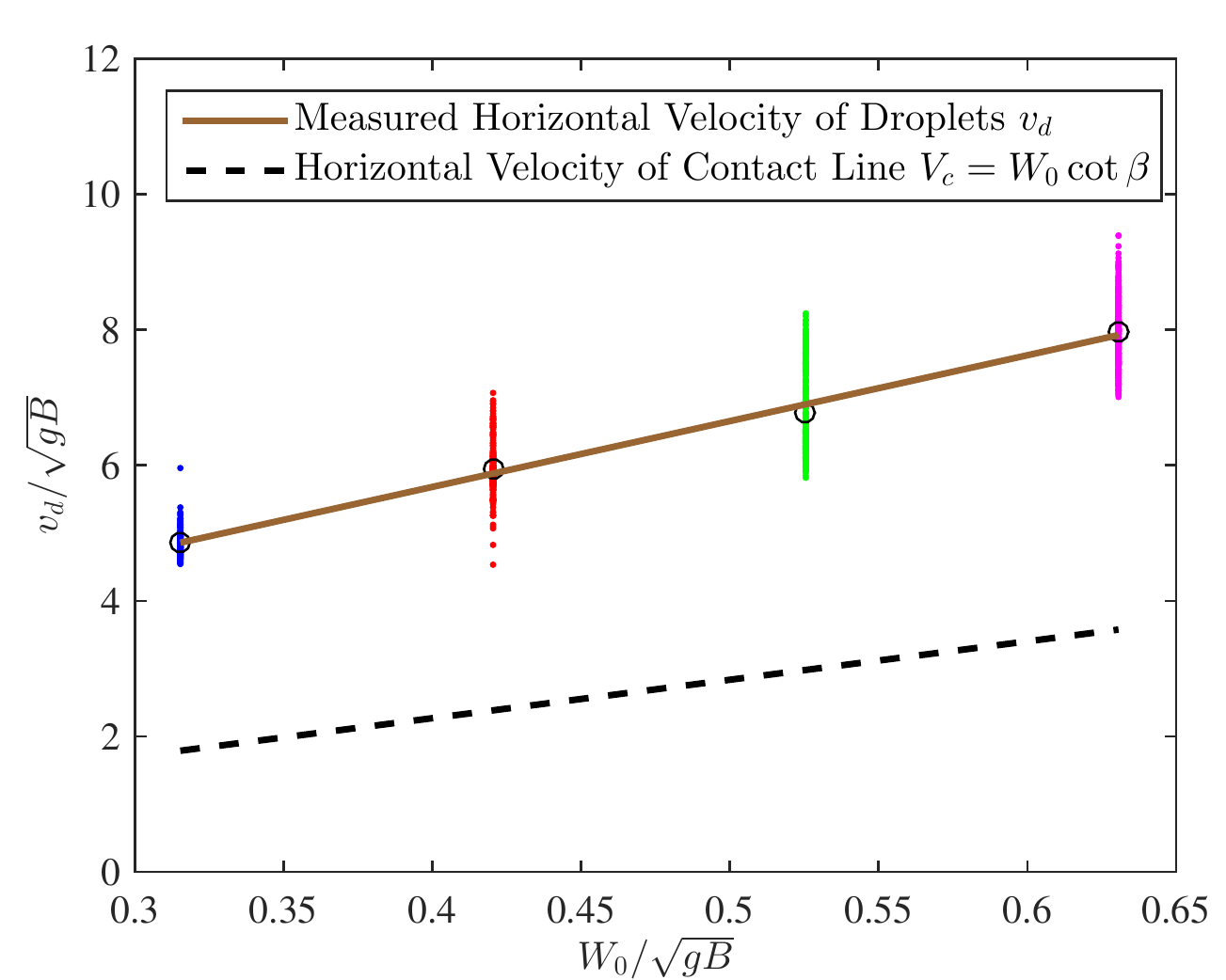}\\
 (b)\\
  \includegraphics[trim={0in 5.5in 1.2in 3.3in},clip = true,width=0.95\linewidth]{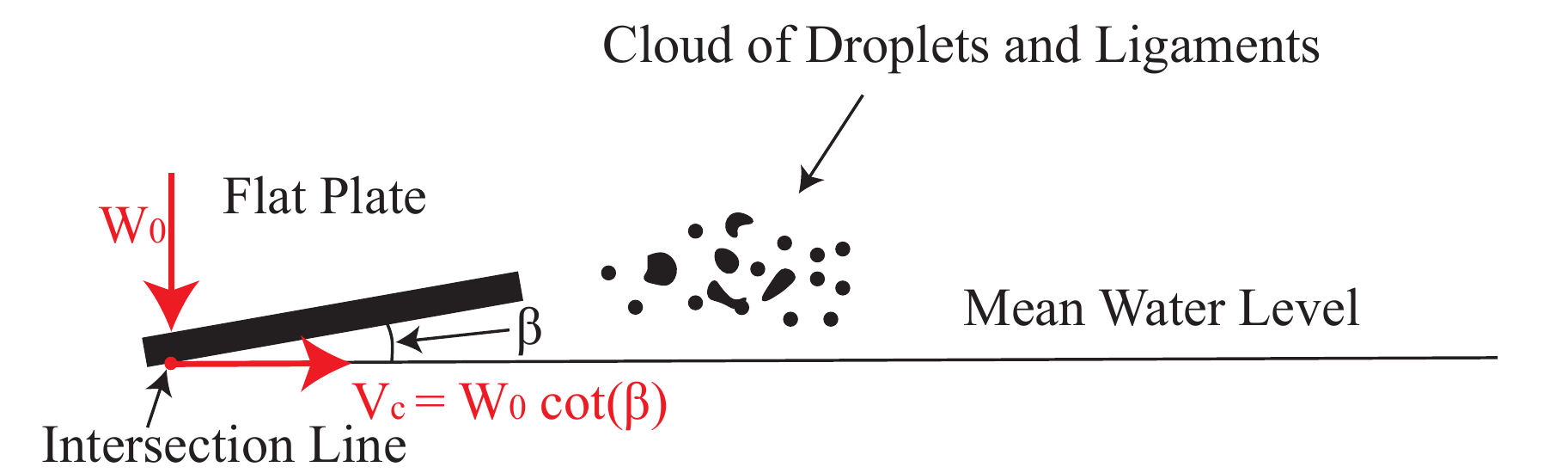}\\
 (c)\\
 \includegraphics[trim={0in 0in 0in 0.2in},clip = true,width=0.95\linewidth]{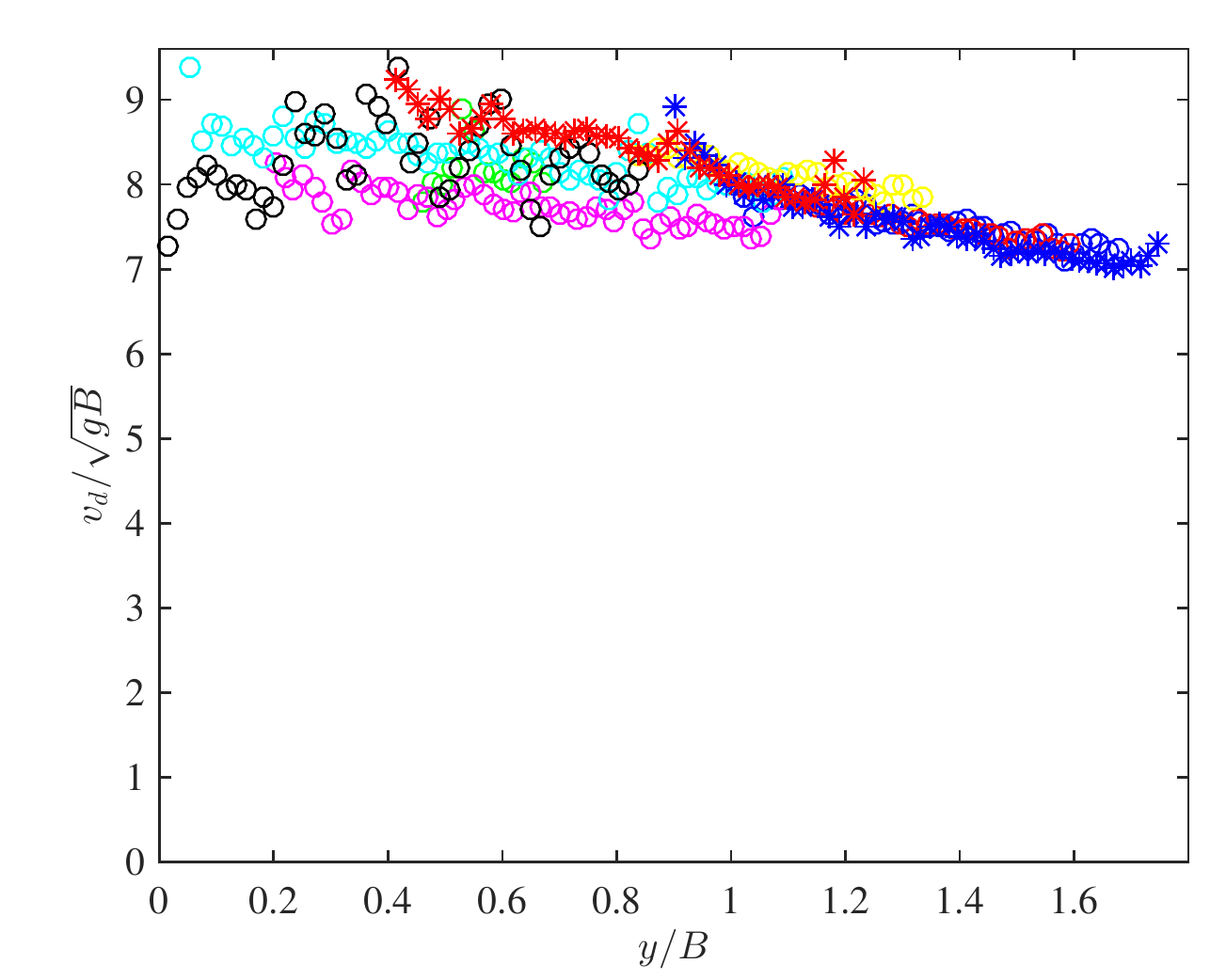}\\
\end{tabular}
\end{center}
\vspace*{-0.25in}
  \caption{Velocity of the droplets in the Type I spray. (a) Velocity of the droplets vs Froude number ($Fr = W_0/\sqrt{gB}$).  The dots with the same color represent the velocities of different individual droplets at different cross-stream positions at the same Froude number. The dashed line is the velocity $V_c (=W_0\cot{\beta})$ of the geometrical intersection point of the plate with the mean water level vs Froude number. The solid straight line is a least squares fit  to the average velocity of the droplets at each Froude number.  (b) The definition sketch for the determination of $V_c$.  (c) The velocity of the droplets vs cross-stream position at $Fr=0.63$. \label{fig:droplet}}
\end{figure}

The velocity ($v_d$) of the droplets in the Type I spray were measured from the LIF image sequences and the resulting data for each of the four values of $Fr$ are plotted  in Figure~\ref{fig:droplet}(a).  The velocities of the droplets are computed based on the change in position of individual droplets from frame to frame in the movies as the droplets move within the plane of the light sheet.  A given data point is the velocity of a droplet at a given cross-stream position ($y$). The spread in the droplet velocities for each Froude number is the result of the difficulty in determining the position of the droplets and the numerical error when taking derivatives of position vs time data, as well as the fact that the droplets slow down as they move away from the plate.  The solid line is a linear fit to the averaged velocity of the droplets at each value of  $Fr$.   The dashed line in the plot is the velocity $V_c (=W_0\cot\beta$) of the geometric intersection point of the plane of the plate and a fixed horizontal line as the plate moves downward at constant speed $W_0$, see Figure~\ref{fig:droplet}(b).  As can be seen in  Figure~\ref{fig:droplet}(a),  the droplet velocity $v_d$ increases as the $Fr$ increases. $v_d$ is more than twice the value of the horizontal velocity of the intersection line at the same $Fr$. The difference between $v_d$ and the horizontal velocity of the geometrical intersection line also increases with $Fr$. 
A plot of droplet velocity versus $y/B$ for $Fr=0.63$ is given in Figure~\ref{fig:droplet}(c).  The data shows quantitatively the slow down of $v_d$ with increasing $y$, probably due to the effect of the drag of the air on the droplets.


\begin{figure*}[!htb]
\centering
  \makebox[0.45\textwidth]{(a)} \makebox[0.45\textwidth]{(b)}
  \includegraphics[width=0.45\textwidth]{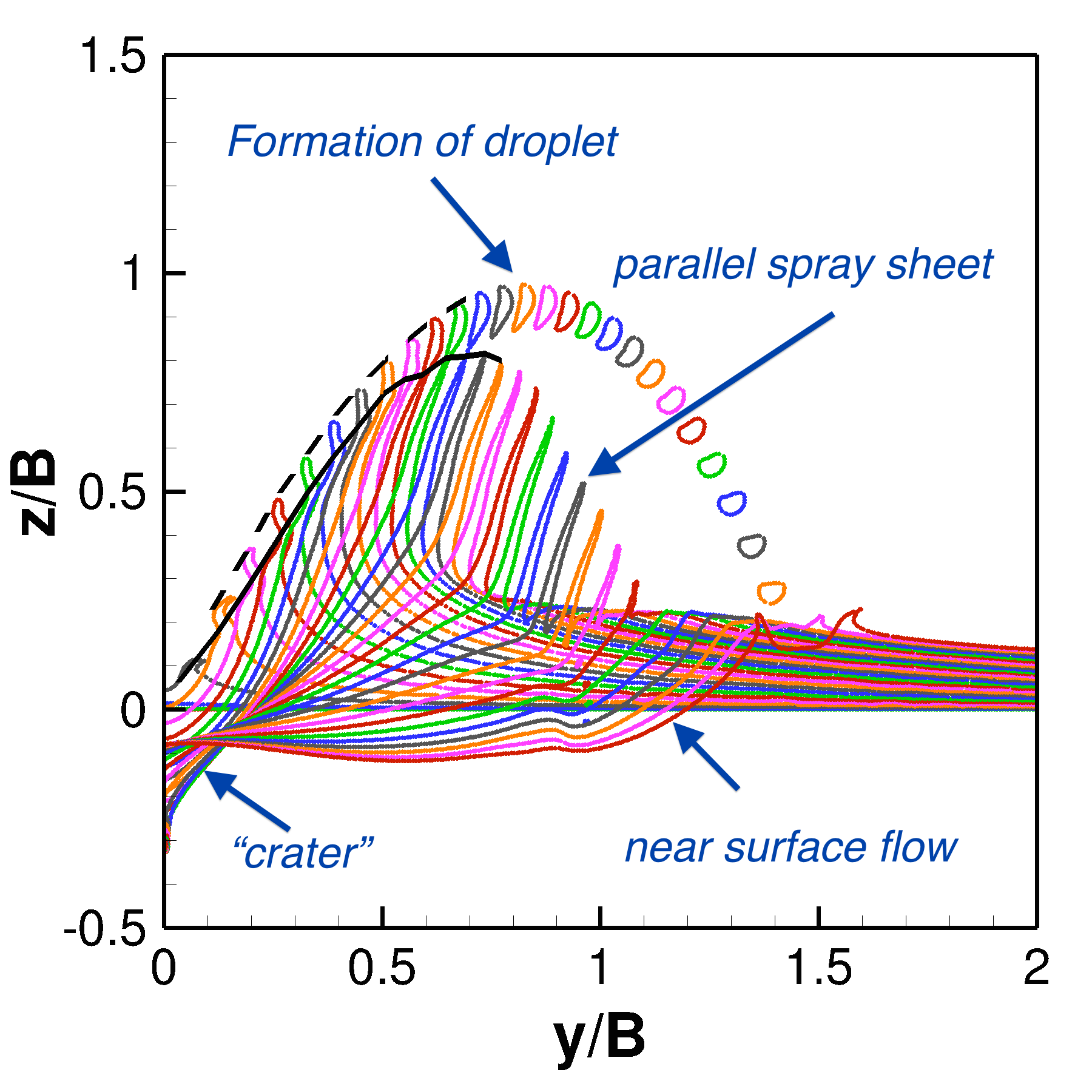}
  \includegraphics[width=0.45\textwidth]{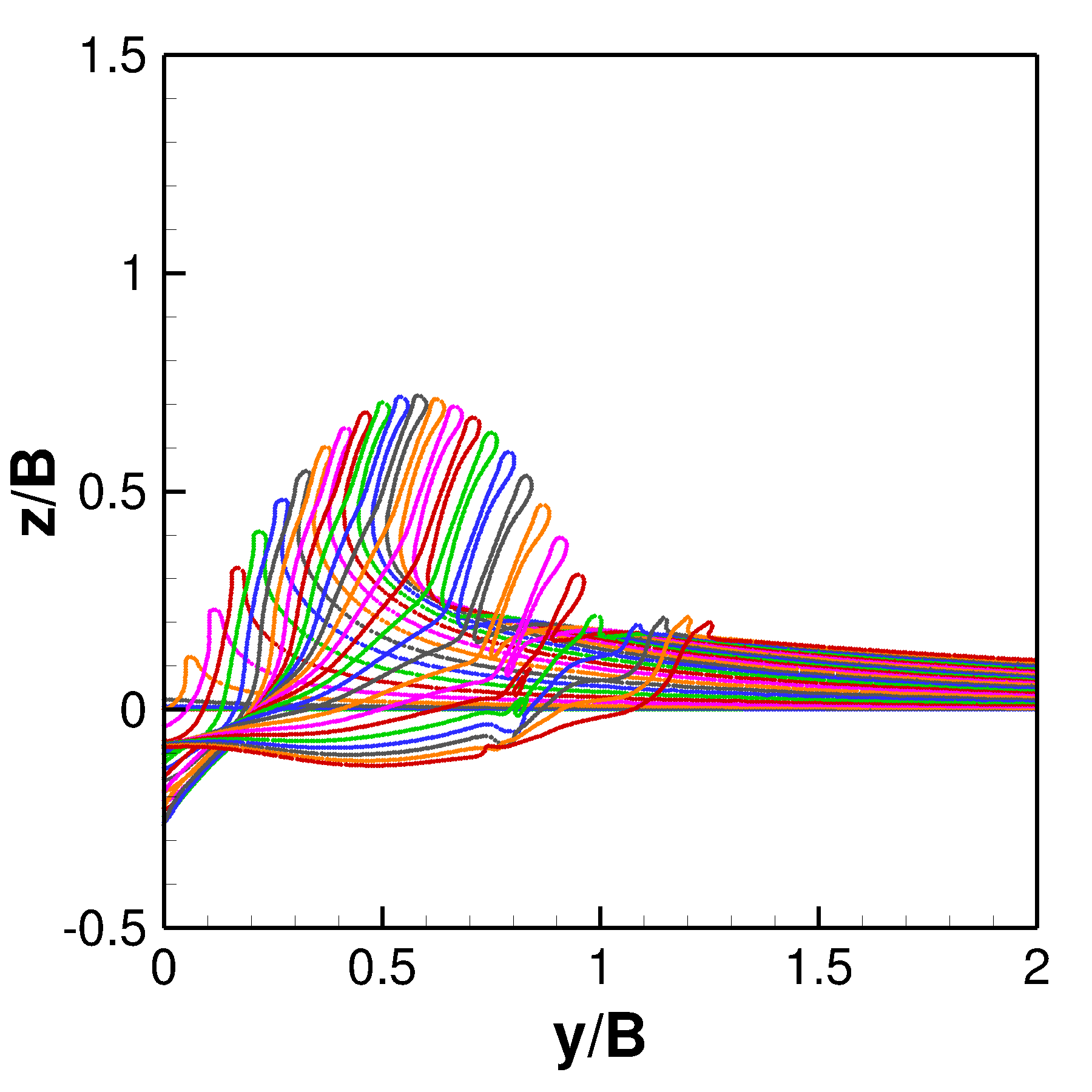}
  \caption{The typical spray profiles in the numerical simulations,
    Fr=0.42, (a) rigid plate; (b) flexible plate with $(EI)^* =
    0.08$. }
\label{fig:spray}  
\end{figure*}
\begin{figure*}[!htb]
\centering
 \makebox[0.45\textwidth]{(a)} \makebox[0.45\textwidth]{(b)}
  \includegraphics[width=0.45\textwidth]{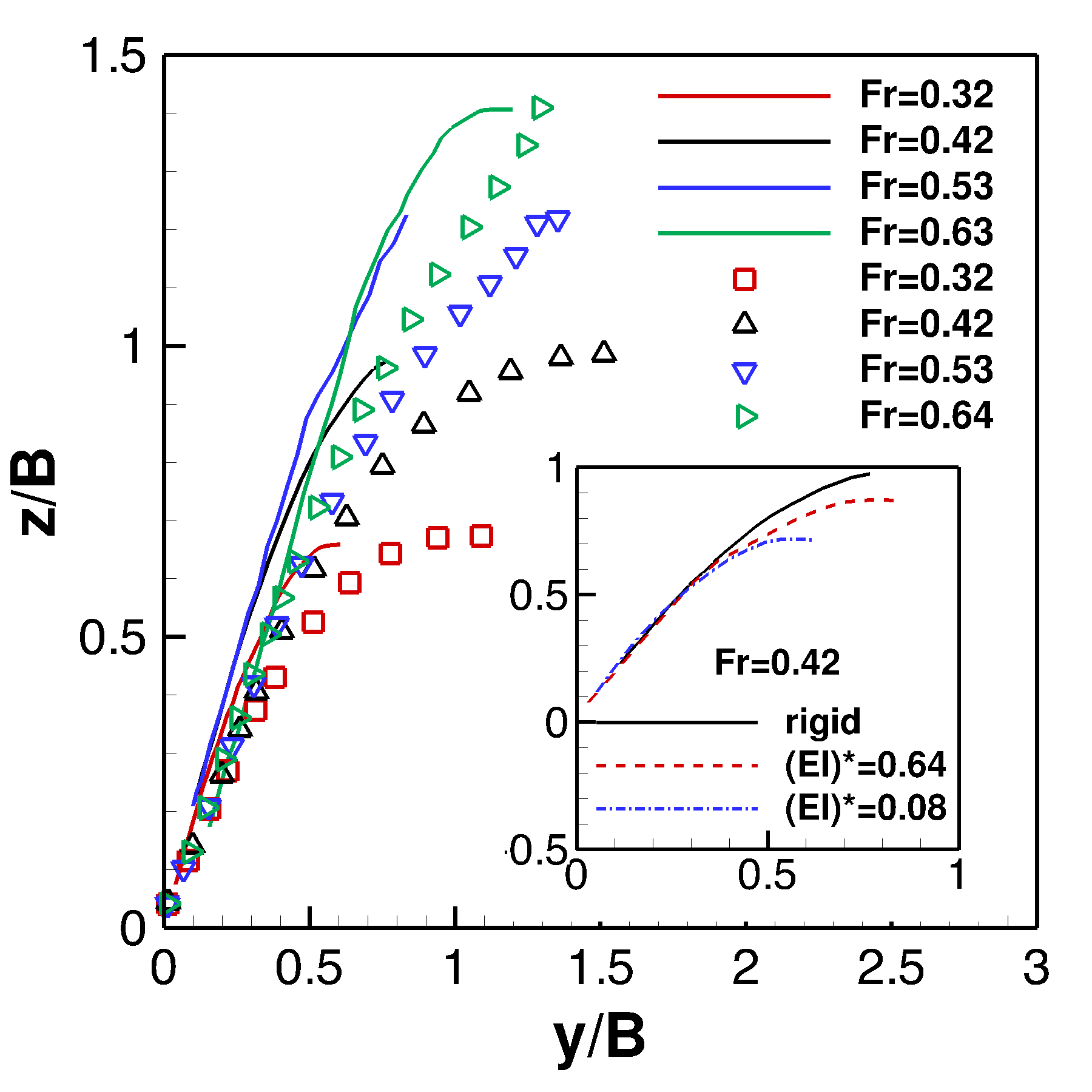}
  \includegraphics[width=0.45\textwidth]{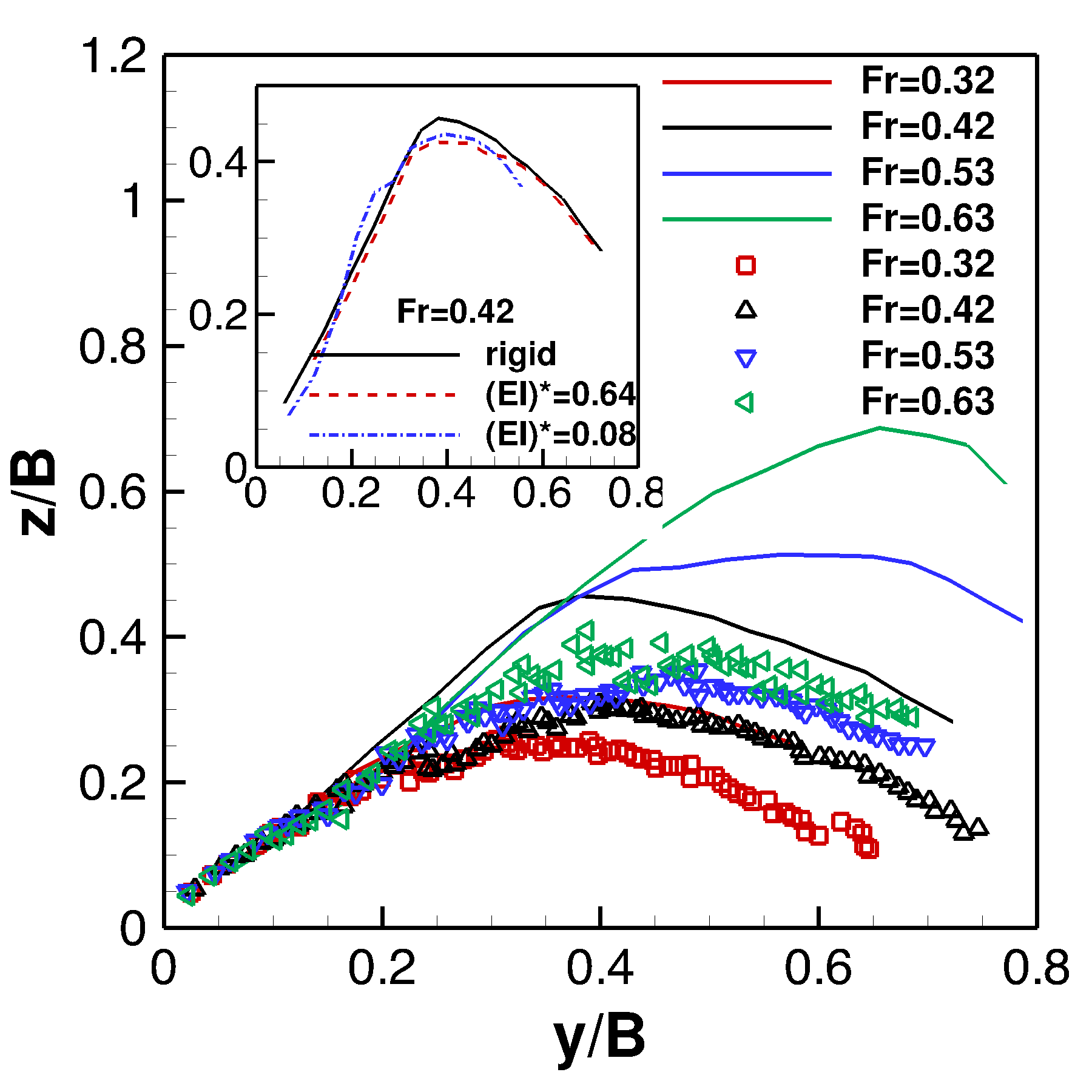}
  \caption{(a) The envelops of the spray profiles and, (b) roots at
    different Froude numbers. The lines show the results from
    numerical simulations, and the symbols show the results from
    experiments. }
\label{fig:envelop}  
\end{figure*}
\begin{figure*}[!htb]
\centering
  \makebox[0.45\textwidth]{(a)} \makebox[0.45\textwidth]{(b)}
  \includegraphics[width=0.45\textwidth]{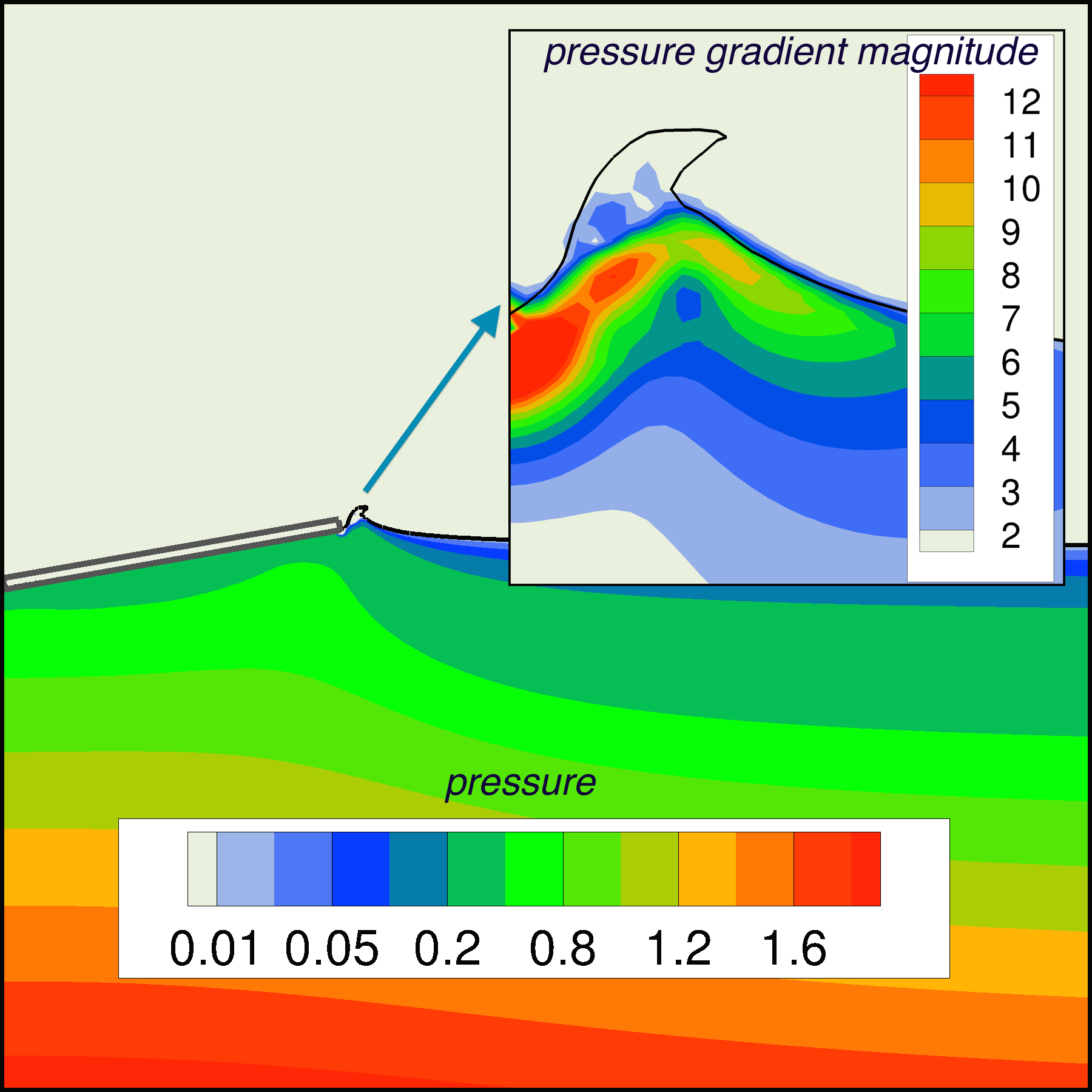}
  \includegraphics[width=0.45\textwidth]{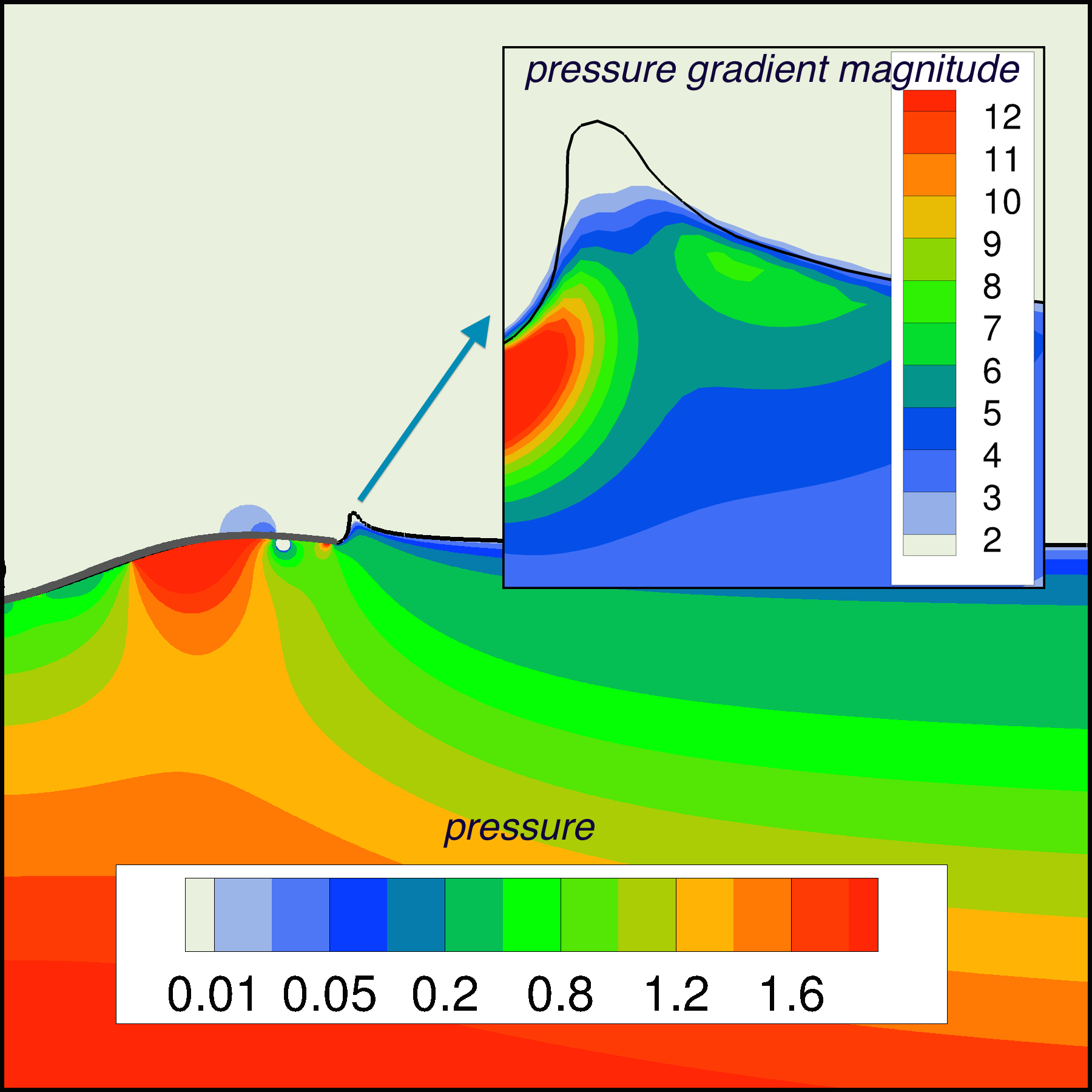}
  \caption{The pressure and pressure gradient magnitude in the case of
    $Fr=0.42$, (a) rigid plate; (b) flexible plate with $(EI)^* =
    0.08$. }
\label{fig:pres}  
\end{figure*}

\subsection{Two-phase Viscous Computations}  \label{sec:results}

The two-dimensional simulations have captured the main features of the
TYPE II spay in the experiments: a wave crest forms at the plate tip,
and propagates to the right; a thin spray sheet is formed from the
wave crest and grows; the spray breaks at the end, and forms droplets,
as shown in Figure~\ref{fig:spray}(a).  The spray root and the profile
envelopes at different Froude numbers have nearly the same slope in the
early stage before they separate from each other, which is consistent
with the experimental results. A direct comparison to the experiments
is shown in Figure~\ref{fig:envelop}.  It can be seen that the
two-dimensional simulations over-predict the height of spray roots and
the envelopes of the spray profiles, and under-predict the distance.
This is an indication that the fluid particles in the computations are
ejected from the wave crest at a higher angle than that in the
experiments. The higher injection angle is probably caused by the 2D
constraint in the numerical simulations, which do not allow for
momentum transfer in the spanwise direction through the spanwise
instabilities that naturally arise in a three-dimensional
configuration.  Despite the quantitative differences between the
numerical and experimental results, the numerical simulation captures
all the main features observed in the experiments, such as the
``crater'', parallel spray sheet, droplet formation, and the disturbed
near surface flow. In addition, the results show the same qualitative
behavior to the changes in the Froude number.

To evaluate the effects of the flexibility of the plate on the results
we considered two cases, where the flexibility of the plate is
varied. The deformation of the plate upon impacting the free surface
varied from 1\% to 7\% as we changed the flexural rigidity from
$(EI)^*=0.64$ to $(EI)^*=0.08$.  The evolution of the spray for one of
the cases with the flexible plate is shown in
Figure~\ref{fig:spray}(b). The overall dynamics are similar to the rigid
plate case. However, the flexibility reduces the height of the spray
and suppresses the formation of droplets.  In particular, the
flexibility mainly affects the evolution of the spray sheet, while the
spray root is relatively less affected, as shown in the insets of
Figure~\ref{fig:envelop}. To better understand the differences in the
spray formation one can examine the pressure gradient near the wave
crest for the rigid and flexible plate cases. Figure~\ref{fig:pres}
shows the distribution of the pressure and pressure gradient magnitude
for $Fr=0.42$. The deformation of the flexible plate changes the
distribution of the pressure and reduces the pressure gradient,
leading to lower spray.

\subsection{ILES Computation}

\begin{figure*}[!htb]
\begin{center}
\begin{tabular}{cc}
(a)&(b)\\
\includegraphics[width=0.43\linewidth]{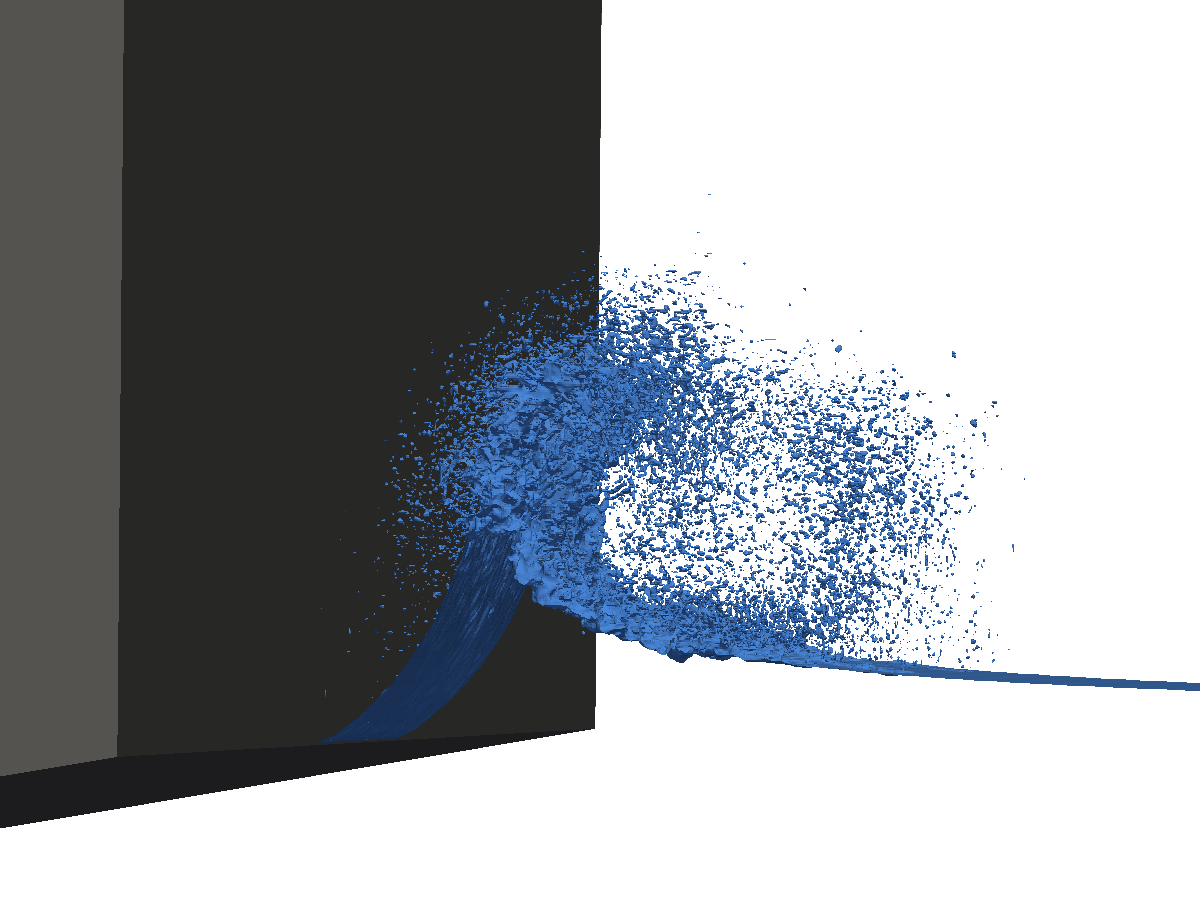}&\includegraphics[width=0.43\linewidth]{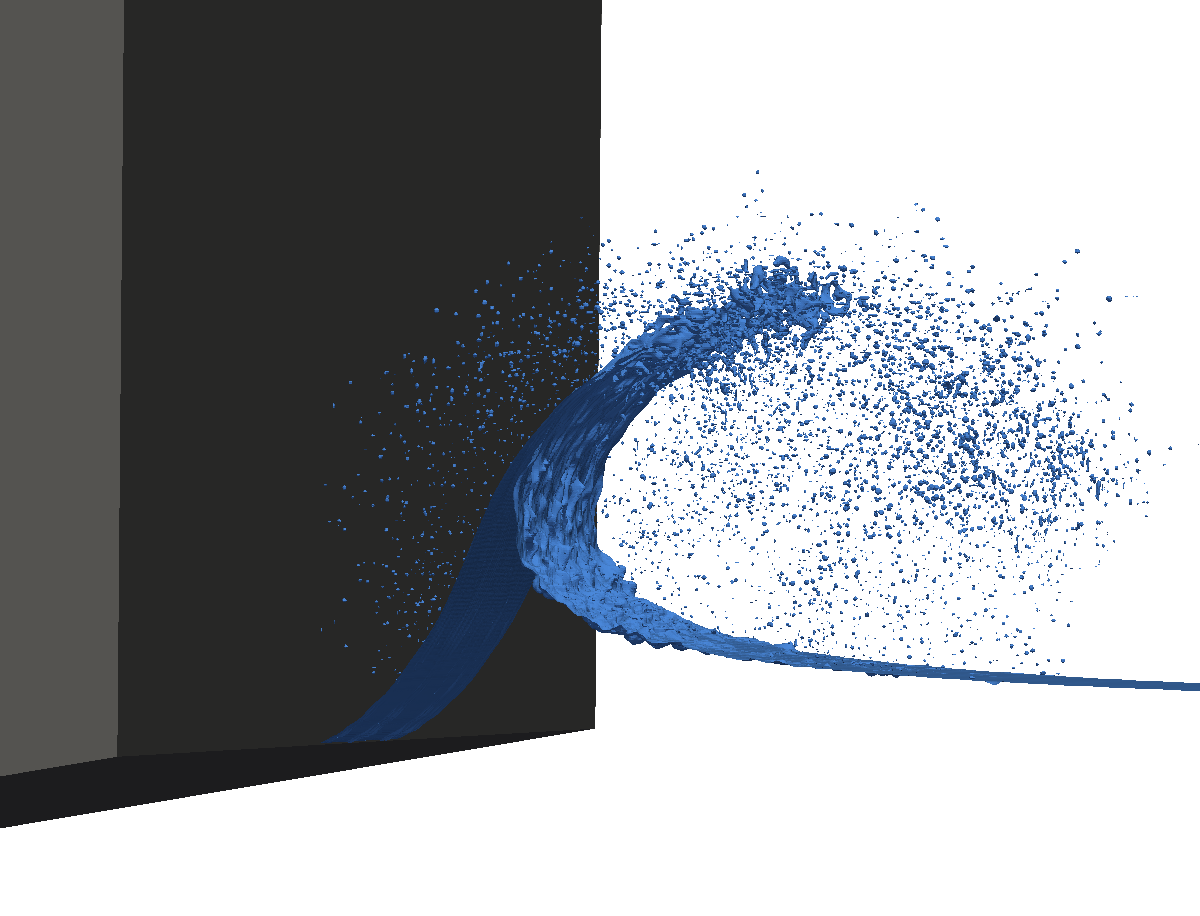}
\end{tabular}
\end{center}
\vspace{-0.25in}
  \caption{NFA simulations at 0.56 seconds. (a) Smoothing every 200 time steps. (b) Smoothing every 40 time steps.\label{fig:NFAsmooth_comp}}
\end{figure*}

\begin{figure*}[!htb]
\begin{center}
\begin{tabular}{cc}
(a)&(b)\\
\includegraphics[width=0.49\linewidth]{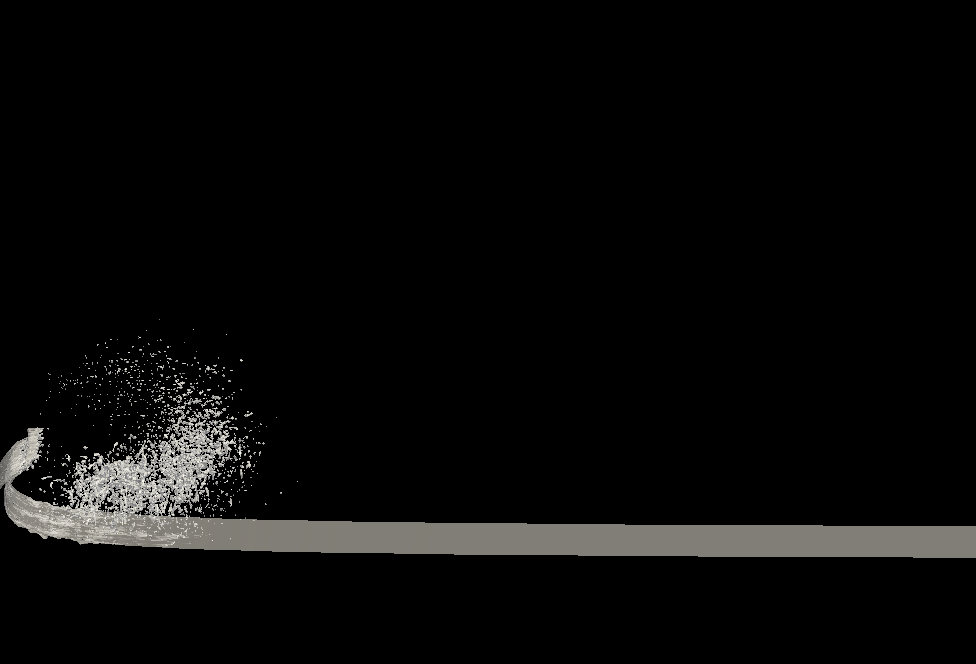}& \includegraphics[width=0.49\linewidth]{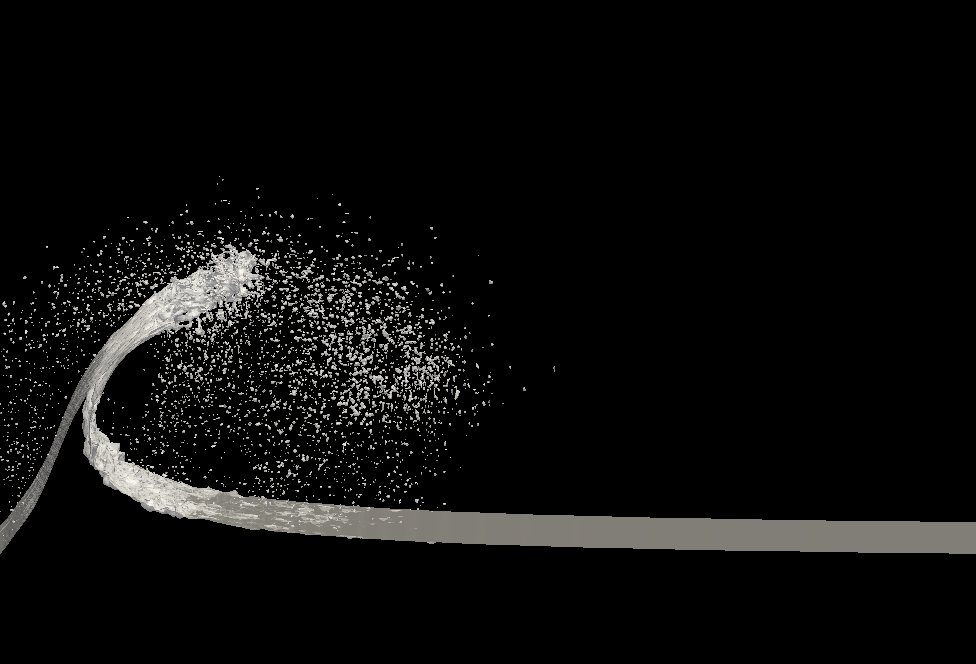}\\
(c)&(d)\\
  \includegraphics[width=0.49\linewidth]{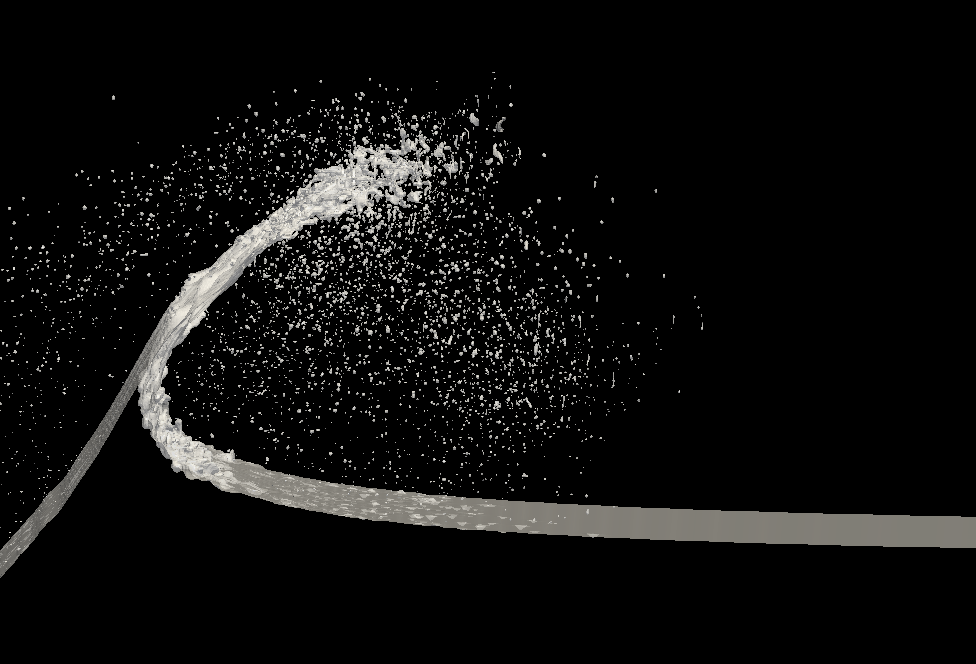}& \includegraphics[width=0.49\linewidth]{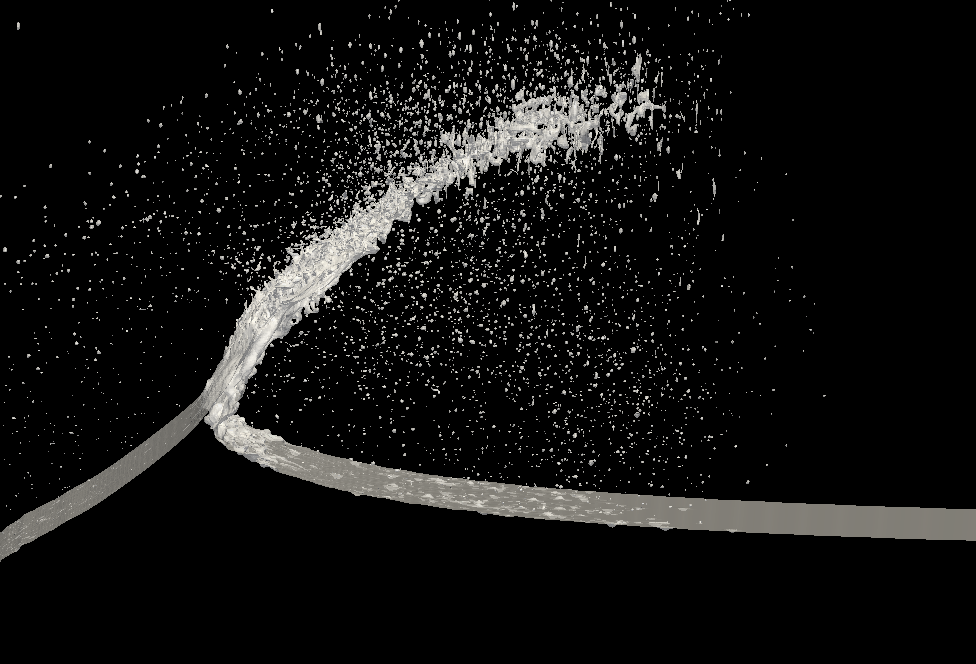}\\
  (e)&(f)\\
   \includegraphics[width=0.49\linewidth]{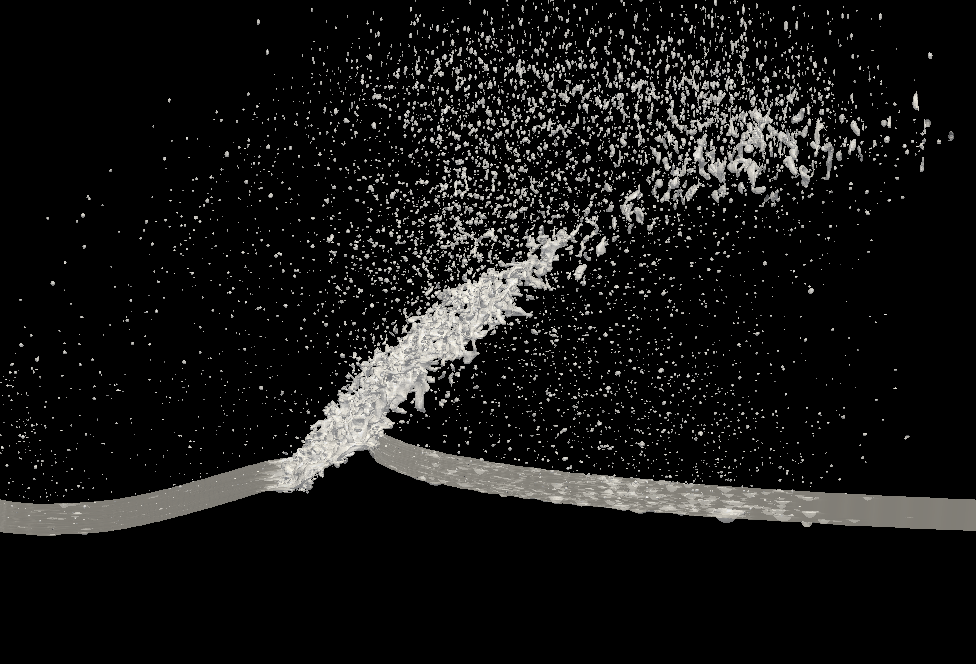}& \includegraphics[width=0.49\linewidth]{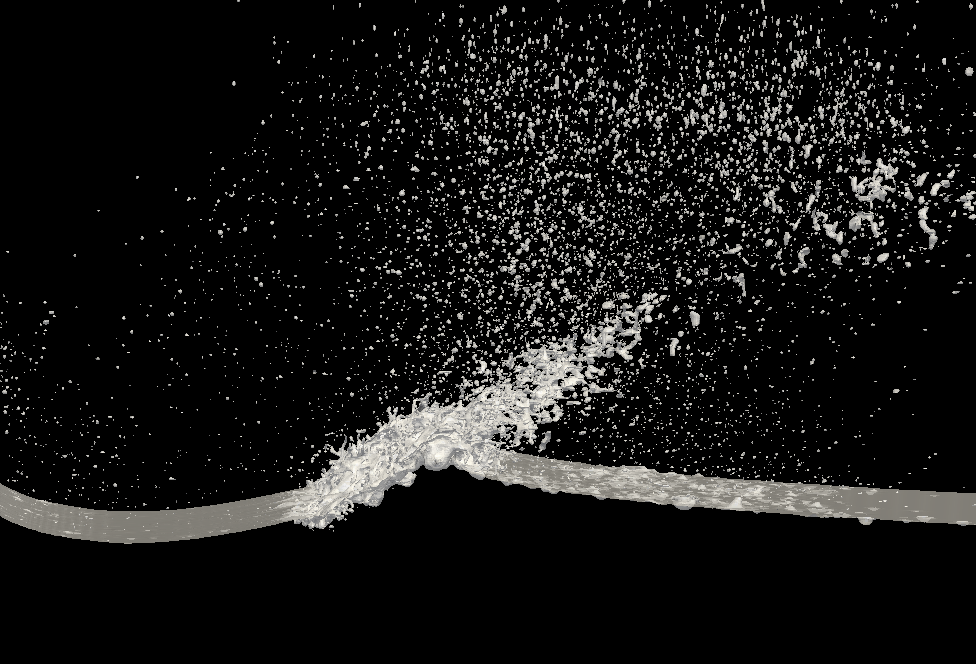}\\
\end{tabular}
\end{center}
  \caption{Sequence of NFA images of the spray formation for $Fr=0.32$.\label{fig:NFAvid_stills}}
\end{figure*}

The initial NFA simulation of the 24 inch per second impact velocity case employed density-weighted velocity smoothing at an interval of every 200 time steps. This corresponds to 0.005 units of nondimensional time. Previous comparisons have shown this amount of smoothing to adequately account for the lack of surface tension and a boundary layer in the air at the interface. In \cite{Drazen:2010} the turbulent breakup of the free surface was found to follow -5/3 power law behavior. Conversely in this simulation of such a high energy and short time scale impact, the breakup of the free surface during the formation of the spray sheet was excessive and prevented the spray sheet from forming. A subsequent simulation used smoothing every 40 time steps. Additional smoothing helped prevent spurious breakup of the free surface into spray and allowed the spray sheet to stay intact for longer. A comparison between the two simulations is shown in Figure \ref{fig:NFAsmooth_comp}. A surface tension model is being developed in NFA so that some of the qualitative nature of this smoothing application will be removed.
 
The simulation with adequate smoothing was able to capture both spray of Type I and Type II. The initial high velocity spray cloud can be seen being ejected from the trailing edge of the plate and travels a short distance across the tank. See image (a) in Figure \ref{fig:NFAvid_stills}. The cloud does not maintain its velocity across the entire width as it does in the experiments. This could be due to the velocity smoothing used to keep the free surface together, or quasi-2D nature of the NFA simulation domain. A simulation with a thick plate, instead of a rectangular prism that goes through the top of the domain would be interesting to run. The width could be increased past the edge of the plate to eliminate any 2D effects. Once the Type I spray is formed, the Type II spray and associated spray root form just as is observed in the experiments. The spray sheet grows in length and breaks up into spray towards the leading edge. See images (b), (c) and (d) in Figure \ref{fig:NFAvid_stills}. The breakup is more vigorous than the experiment but the overall features are qualitatively similar. Additional smoothing would most likely keep the spray sheet from breaking up and would be more consistent with the experimental observations. The spray sheet falls under the effect of gravity and forms a free surface turbulent splash zone as seen in the experiments. See images (e) and (f) in Figure \ref{fig:NFAvid_stills}. The crater that is formed in the experiment behind the spray root is also seen in the simulation. The images in Figure \ref{fig:NFAvid_stills} are taken from a video of the simulation with a field of view that is similar to the experimental images in Figure \ref{fig:spray_pic}.

\section{Conclusions}
The spray generated by the impact of a flat plate with roll angle $\beta=10^\circ$ on a quiescent water surface is studied by experiments,  viscous flow simulations and ILES flow simulations at different impact Froude numbers. The formation of two types of sprays, called Type I and Type II, is found in both the experiments and the simulations. The Type I spray is a cloud of high-speed droplets and ligaments generated when the leading (lower) edge of the plate impacts water surface. The Type II spray is formed after the trailing (upper) edge of the plate starts to impact on the local water surface. Type II spray consists of a growing crater, a thin spray sheet and droplets that break up from the sheet. Both numerical simulations exhibit spray of this type.

The surface profiles of the Type II spray are measured and computed and several geometrical parameters are studied. The behaviors of the spray envelope and the spray root point trajectory are found to be independent of Froude number close to the trailing edge of the plate, while further away from the trailing edge, the envelopes and root point trajectories reaches higher vertical positions for larger Froude numbers. The numerical simulations captures the general behavior of the Type II spray and agrees favorably with the experimental results.

The velocity ($v_d$) of the droplets in the Type I spray is measured in the experiments. It is found that, at the same $Fr$, $v_d$ is larger than the horizontal velocity of the geometrical intersection point of the plate of the plate and the plane of the mean water level. As the droplets move farther from the edge of the plate, $v_d$ decreases, probably due to the effect of the drag of the air.

\subsection{Acknowledgements}
The support from Office of Naval Research under grants N000141310587 (Program manager: Dr.\ Steven Russell)  is gratefully acknowledged. 

\bibliography{29ONR,29ONR_2}
\bibliographystyle{29ONR}

\end{document}